\documentclass[aps,amsmath,twocolumn,amssymb,floatfix,showpacs,superscriptaddress,nofootinbib,longbibliography]{revtex4-1}
\usepackage{mathtools}
\usepackage{braket}
\usepackage[dvipsnames]{xcolor}
\usepackage{float}
\usepackage{subfigure}
\usepackage{graphics}
\usepackage{bm}
\usepackage{soul}
\usepackage{MnSymbol}
\usepackage{braket}
\usepackage{tikz}
\usepackage{blkarray}
\usepackage{pgffor}
\usepackage{multirow}
\usepackage{booktabs}
\usepackage{silence}
\usepackage{makecell}
\usepackage[colorlinks=true,linktoc=page,linkcolor=BrickRed,citecolor=magenta,urlcolor=purple]{hyperref}

\mathchardef\mhyphen="2D 

\newcommand\bea{\begin{eqnarray}}
\newcommand\eea{\end{eqnarray}}
\newcommand\beq{\begin{equation}}  
\newcommand\eeq{\end{equation}}

\usepackage[normalem]{ulem}
\definecolor{lime}{HTML}{A6CE39}
\usepackage{sidecap,tikz}
\DeclareRobustCommand{\orcidicon}{\hspace{-1.0mm}
	\begin{tikzpicture}
		\draw[lime, fill=lime] (0.0,0.0) 
		circle [radius=0.15] 
		node[white] {{\fontfamily{qag}\selectfont \tiny \,ID}};
		\draw[white, fill=white] (-0.0525,0.095) 
		circle [radius=0.007];
	\end{tikzpicture}
	\hspace{-3.0mm}
}
\foreach \x in {A, ..., Z}{\expandafter\xdef\csname orcid\x\endcsname{\noexpand\href{https://orcid.org/\csname orcidauthor\x\endcsname}
		{\noexpand\orcidicon}}
}

\AtBeginDocument{%
	\newwrite\bibnotes
	\def\bibnotesext{Notes.bib}
	\immediate\openout\bibnotes=\jobname\bibnotesext
	\immediate\write\bibnotes{@CONTROL{REVTEX41Control}}
	\immediate\write\bibnotes{@CONTROL{%
			apsrev41Control,author="08",editor="1",pages="1",title="1",year="1"}}
	\if@filesw
	\immediate\write\@auxout{\string\citation{apsrev41Control}}%
	\fi
}%
\begin{document}

\title{Superconducting order parameter in aperiodic binary systems}
\author{Sougata Biswas\orcidA{}}
\email{sougatabiswas@prl.res.in}
\affiliation{Theoretical Physics Division, Physical Research Laboratory, Navrangpura, Ahmedabad-380009, India}

\author{Debika Debnath\orcidB{}}
\email{debika.uoh@gmail.com}
\affiliation{Theoretical Physics Division, Physical Research Laboratory, Navrangpura, Ahmedabad-380009, India}

\author{Paramita Dutta\orcidC{}}
\email{paramita@prl.res.in}
\affiliation{Theoretical Physics Division, Physical Research Laboratory, Navrangpura, Ahmedabad-380009, India}

\vspace{0.8cm}
\begin{abstract}
Recent discovery of the superconducting ground state in systems lacking perfect periodicity but with long-range ordering has opened up an exciting new avenue for superconductivity based on aperiodic systems. In this work, we explore the scope by theoretically investigating the behavior of the superconducting order parameter (OP) in aperiodic binary systems (ABSs), both Fibonacci and non-Fibonacci type, based on the attractive Hubbard model. We begin with one-dimensional toy model, and for the generality of our findings, we extend our analysis to two-dimensional ABSs. Remarkably, despite the increased dimensionality, the qualitative features of the OP remain largely preserved. By systematically analyzing models generated through various growth rules, we elucidate the influence of aperiodicity on the OP amplitudes and how it evolves towards the periodic limit with the change in the structural pattern. We study the evolution of the OP with respect to the temperature, strength of the interaction, and nearest-neighbor hopping amplitude. Our numerical analysis identifies the most favorable ABS and parameter regime that support enhanced onsite pairing amplitudes. Additionally, we provide a comparative analysis of the superconducting transition temperatures across the range of aperiodic configurations. To gain further insight into these systems, we compute key thermodynamic quantities: the entropy and electronic specific heat, and examine their dependence on the underlying structural sequences. This analysis enables us to determine which ABSs are most conducive to Cooper pair formation. 
\end{abstract}

\maketitle

\section{Introduction}
\label{intro}
The discovery of metallic phases in materials that lack translational symmetry but exhibit long-range orientational order, known as quasicrystals (QCs), has opened an exciting new frontier in condensed matter physics research in 1984~\cite{Shechtman1984}. Although QCs do not possess the periodicity found in regular crystals, their atomic arrangements follow deterministic rules that establish long-range order into its configuration~\cite{Levine1984, Kohmoto1987, Liu1987}. In regular crystals with perfect periodicity, electron wavefunctions typically extend throughout the lattice, leading to metallic behavior, whereas in disordered systems, wavefunctions become mostly localized, exhibiting an insulating phase. Interestingly, QCs offer a rich landscape for the spectral and wave function's behavior due to the presence of the structural pattern i.e., quasiperiodicity. The eigenstates are neither fully localized nor fully extended in these quasiperiodic systems (QPSs)~\cite{Fujiwara1989, Kohmoto1987, Mace1996, Ryu1993, Liu1987, Niu1986}. These are critical states that feature unique spatial structures characterized by multifractal scaling. QPSs often exhibit energy spectra resembling a Cantor set~\cite{Kohmoto1987, Macia1996, Mace1996, Fu1991, Kroon2002}, offering a testing ground for exploring unusual transport and localization phenomena~\cite{Capaz1990, Jagannathan2021, Mace2017, Macia1999}. QPSs and many other aperiodic systems have been realized experimentally using the fabrication of multilayered structures~\cite{Merlin1985, Elser1985, Todd1986, He1988, Merlin1987, Mayou1993,   Zarate2001, Dal2004, Yamamoto2004, Hendrickson2008, Passias2009, Nagao2015, Mihalkovik2020, Reisner2023, Wuhrl2023}.

The long-range ordering in QPSs manifests in various sequences or patterns depending on the dimension of the system. Among one-dimensional ($1$D) QPSs, Fibonacci quasiperiodicity is the most well-known structure where the number of sites or bonds in successive generations grows according to the Fibonacci binary sequence. Many such aperiodic systems can be generated by extending or generalizing the recurrence relations used for Fibonacci (Golden-mean), giving rise to a zoo of aperiodic binary systems (ABSs)~\cite{Macia2107} like Silver-mean, Bronze-mean, Copper-mean, Nickel-mean etc. Other $1$D binary aperiodic models like Thue-Morse, period-doubling have also emerged as prototypical models for exploring the interplay between aperiodicity and various quantum mechanical properties~\cite{Oh1993, Kolar1991, Cheng1988, Chattopadhyay2001, Carpena1995, Zarate2001, Velasco2002, Zheng1987, Niu1990, Piechon1995, Chakrabarti2003, Mordret2025, Schirmann2024, Dutta2014, Das1988, Rontgen2019, Matsubara2025}. These systems have been extensively investigated, particularly in relation to their spectral characteristics, thermodynamic properties, and wave function behavior~\cite{Luck1989, Chakrabarti1991, Sil1993, Ryu1992, Ghosh1998, Zheng1987, Gumbs1995, Nori1986, Kohmoto1986, Lambropoulos2019}. In addition to static and spectral characteristics, significant attention has been devoted to understanding the quantum dynamics in ABSs~\cite{Dulea1992, Katsanos1995, Brito1995, Enrique1996, Piechon1996, Ketzmerick1997, Vidal1999, Vidal2001, Chiaracane2021, Stefanie2009, Sougata2024}, enriching the study on these systems as a bridge between regular crystals and disordered materials.

Recent works on QCs have renewed the interest of the community. Topological phases have been shown to appear in QCs, extending the concept beyond the conventional paradigm of metal-insulator phases~\cite{Kraus2012, Longhi2019, Lin2022, Chen2019, Chen2023topo, Rangi2024}. The possibility of boundary states in the topological phases makes them an important platform for exploring quantum phases of matter~\cite{Baboux2017, Peng2024, Chen2020, Huang2019, Mor2013, Ouyang2024, Duncan2020, Chen2024b, Wang2022}. On top of that, the recent discovery of superconducting ground state in Fibonacci QPSs and their approximants based on the Al–Zn–Mg alloy has marked a possibility of QPS or in general ABS-based superconductivity~\cite{kamiya2018}. This recent discovery highlighted the relevance of quasiperiodic structures for realizing superconducting phases. Fibonacci QPSs have been taken as the testbed to study the superconductivity, as it is one of the simplest $1$D quasiperiodic models, which can provide a valuable framework for exploring fundamental physical phenomena. The superconducting properties of Fibonacci QPSs have been explored to understand how the absence of translational symmetry influences superconducting pairing~\cite{Sun2024, Wang2024, Sandberg2024, Kobialka2024, Rai2019, Rai2020b}. It has also been studied in $1$D Aubry-Andr\'e-Harper~\cite{R7, R8} model and their two-dimensional ($2$D) and three-dimensional ($3$D) extensions~\cite{Wang2024, R7}, Penrose and Ammann-Beenker tilings~\cite{Rasoul2021, Rolando2019, Shiro2017, Liu2025, Manna2024}, and also in Moir\'e materials~\cite{R9}. Superconductivity has been realized in dodecagonal~\cite{R1, R2} and decagonal~\cite{R3} systems and also in their approximants ~\cite{kamiya2018, R4, R5}, with the quasiperiodicity either in full $3$D~\cite{kamiya2018} or in $2$D layers with periodic stacking~\cite{R3}, and in Moir\'e materials~\cite{R6}.
Superconductivity induced by phason has also been reported in incommensurate structures~\cite{Jiang2023}, where phason is a quasiperiodic degree of freedom similar to phononic degrees of freedom in periodic crystals. These studies have revealed that aperiodic order can give rise to both conventional and unconventional superconducting phases where the superconducting gap and the local pairing amplitude exhibit non-uniform, often fractal-like spatial profiles. It also includes the appearance of topological superconductivity~\cite{Rasoul2021, Kobialka2024, Hori2024}. 

The emergence of superconductivity and the corresponding behavior of the pairing amplitude have been extensively investigated across a wide variety of periodic systems, including those subjected to various forms of disorder~\cite{David1958, Amit2001, Tanaka2000, Liyu2023, Amit2000, Croitoru2000, Bai2023a, Bai2023b, Chen2023, Chen2024c, Debmalya2017, Wang2023, Aryanpour2006}. However, a natural question appears here: what happens in other QPSs with different structures? Given that different quasiperiodic structures have different construction patterns, do their superconducting properties also vary accordingly? Not only QPSs, the same question can be generalized for ABSs. How to find the best candidate for the Cooper pair formation within the zoo of ABSs? How do they approach periodic limit? Motivated by these fundamental questions, we present a detailed study on the superconducting order parameter (OP) in a variety of ABSs including QPSs. We start with $1$D toy models and then extend our analysis to $2$D ABSs to examine the validity of the mean-field approach. As a prototype, we consider Fibonacci (Golden-mean), Silver-mean, and Bronze-mean QPS from call them as ABS-I. Keeping in mind the sequence generation of ABS-I, we also consider Copper-mean, Nickel-mean, Thure-Morse and Period-doubling as other members of ABSs family. For the sake of clarity and simplicity, we refer to Copper-mean and Nickel-mean ABS as ABS-II as the growth rules are opposite to that of ABS-I as mentioned in Table-\ref{table}, whereas the Thue-Morse and Periodic-doubling ABSs are refereed as ABS-III, as their growth rules are quite different from ABS-I or ABS-II. Note that, both ABS-II and ABS-III differ significantly in their local environments, hierarchical structure, and spectral characteristics from the ABS-I~\cite{macia2008, Macia2107}\footnote{Members of ABS-I are the QPS, and they are generated by increasing $t_A$. In contrast, the members of ABS-II are constructed through a different growth pattern by increasing $t_B$. Meanwhile, the members of ABS-III follow distinct growth pattern. This classification is introduced solely for the sake of clarity and comparison.}.
Important questions arise about the universality or variability of key features, such as the spatial pattern of the OP, the temperature dependence of the OP, and their thermodynamic properties. 

The primary goal of our study is to understand the origin of the behavior of the OP across the range of ABSs we examine. We find that while each aperiodic sequence exhibits spatial variation in the OP, these patterns and their magnitudes are not universal; instead, they strongly depend on the underlying sequence and the range of parameters governing the generation of the sequence. Using an attractive Hubbard model, we investigate how the OP distributions are affected by the aperiodicity. We also provide a comparative analysis to identify the conditions favorable for Cooper pair formation. We show that OP amplitudes are strongly sensitive to the sequences of the  ABSs and also to the model parameters. The transition temperatures vary largely with the configurations and parameter values of the ABSs as well. Our findings offer guidelines for a broad class of aperiodic systems to identify which QPS or ABS is the best choice for higher OP, with potential implications spanning from condensed matter theory to material science, quantum optics, and superconducting device fabrication.

We organize the rest of the article as follows. In Sec.~\ref {II}, we introduce various ABS models and Hamiltonians, and describe the theory employed for the present study. Our results for the spatial distribution of OPs, transition temperatures, and thermodynamic properties of various $1$D models are discussed in Sec.~\ref{III}. Next, we extend our study on $1$D ABSs to $2$D structures and investigate their properties in Sec.~\ref{III}. Finally, we summarize and conclude in Sec.~\ref{IV}.

\section{Model and Hamiltonian}
\label{II}
We start by describing our $1$D ABS model using an attractive Hubbard Hamiltonian which reads
\begin{eqnarray}
    H & = & -\sum_{i,\sigma} (t_{i,i+1} c_{i,\sigma}^{\dagger}c_{i+1,\sigma} + \text{h.c.}) +\sum_{i,\sigma}(\epsilon - \mu)c_{i,\sigma}^{\dagger}c_{i,\sigma}\nonumber\\&&- U\sum_{i} c_{i,\uparrow}^{\dagger}c_{i,\uparrow}c_{i,\downarrow}^{\dagger}c_{i,\downarrow}
    \label{ham}
    \end{eqnarray}
where $c_{i,\sigma}^{\dagger} (c_{i,\sigma})$ operator creates (annihilates) an electron of spin $\sigma = (\uparrow,\downarrow)$ at $i$-${\rm th}$ site, $t_{i,i+1}$ is the nearest-neighbor hopping term which can be $t_{A}$ or $t_{B}$ depending upon the growth rule of the aperiodic binary sequences. The terms $\epsilon$ and $\mu$ are the onsite and chemical potential, respectively. $U$ is the strength of the attractive Hubbard interaction. Throughout the present study, we consider bond models for all ABSs where the hoppings are distributed following various growth rules that determine the arrangement in an aperiodic but long-range ordered fashion. 
\begin{table}
\begin{center}
\caption{Growth Rules of ABSs}
\renewcommand{\arraystretch}{1.2}
\setlength{\tabcolsep}{8pt}
\begin{tabular}{|c|c|c|}
\hline
Class & ABSs & Growth Rules \\
\hline \hline
ABS-I & Fibonacci & $t_A \rightarrow t_A t_B,$  $t_B \rightarrow t_A$ \\
\cline {2-3}
(QPS) & Silver-mean & $t_A \rightarrow t_A t_ At_ B,$  $t_B \rightarrow t_A$ \\
\cline{2-3}
& Bronze-mean & $t_A \rightarrow t_A t_A t_A t_B,$  $t_B \rightarrow t_A$ \\
\hline \hline
ABS-II & Copper-mean & $t_A \rightarrow t_A t_B t_B,$  $t_B \rightarrow t_A$ \\
\cline{2-3}
& Nickel-mean & $t_A \rightarrow t_A t_B t_B t_B,$  $t_B \rightarrow t_A$ \\
\hline \hline
ABS-III & Thue-Morse & $t_A \rightarrow t_A t_B,$  $t_B \rightarrow t_B t_A$ \\
\cline{2-3}
& Period-doubling & $t_A \rightarrow t_A t_B,$  $t_B \rightarrow t_A t_A$ \\
\hline
\end{tabular}
\label{table}
\end{center}
\end{table}

The growth of the generalized Fibonacci sequence follows a rule based on the replacements $t_A \xrightarrow{}t_A^{i}t_B^{j}$ and $t_B \xrightarrow{}t_A$. Now, by a proper choice of $i$ different quasiperiodic sequences are generated with $j=1$. The total number of bonds in $(n+1)$-th generation is given by, $G_{n+1} = i G_{n} + j G_{n-1}$, with $n \ge 1$ and $G_{0} = G_{1} = 1$. Here we generate $1$D QPS by choosing $i=1$ for the Fibonacci (Golden-mean), $i=2$ for the Silver-mean, and $i = 3$ for the Bronze-mean QPS. All these three sequences belong to the family of metallic means characterized by the golden ratio, silver ratio, and the bronze ratio, respectively. We refer them as ABS-I. Keeping $i = 1$, two other aperiodic binary sequences are generated considering $j= 2$ for the Copper-mean, and $j =3$ for the Nickel-mean system~\cite{Macia2006}. We refer to these two ABSs as ABS-II as their growth rules are quite similar. Our investigation is also extended to other well-known ABSs: Thue-Morse and period-doubling, which are generated by growth rules completely different from that of ABS-I or ABS-II. So, we refer these two as ABS-III. The growth rules for all these ABSs are shown in Table-\ref {table}. Their growth rules physically signify different long-range ordering or cluster and the degree of aperiodicity affecting various properties of the lattice.
\begin{figure*}
\centering
\includegraphics[width= 0.8\linewidth]{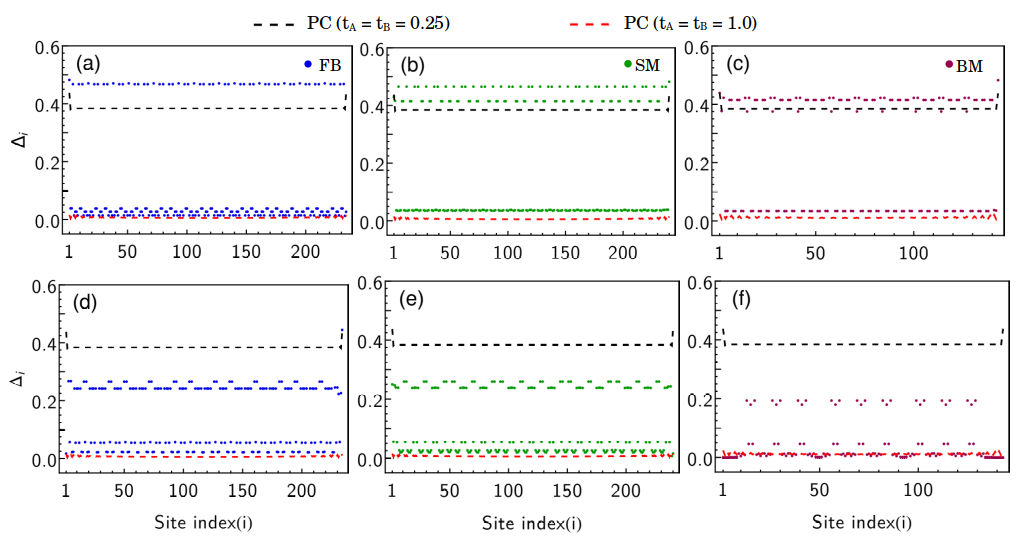}
\caption{Distribution of local OP in (a,d) Fibonacci or Golden-mean (FB), (b,e) Silver-mean (SM), and (c,f) Bronze-mean (BM) QPS for the Hubbard interaction strength $U = 1$, temperature $T = 0.01$, and the hopping amplitudes: (a-c) $t_{A} = 0.25$, $t_{B} = 1$ and (d-f) $t_{A} = 1$, $t_{B} = 0.25$. In each plot, the black and red dotted line represent the OP distribution of a periodic chain (PC) with $t_A = t_B = 0.25$ and $t_A = t_B = 1$, respectively. } 
\label{OI1}
\end{figure*}

We cast our bond model into the Bogoliubov–de Gennes (BdG) Hamiltonian within the mean field approximation~\cite{de2018superconductivity}. The resulting Hamiltonian is given by,
\begin{eqnarray}
      H_\text{MF} & = &-\sum_{i,\sigma} (t_{i,i+1} c_{i,\sigma}^{\dagger}c_{i+1,\sigma} + \text{h.c.}) +\sum_{i,\sigma}(\epsilon - \mu)c_{i,\sigma}^{\dagger}c_{i,\sigma}\nonumber\\ && + \sum_{i}(\Delta_{i}c_{i\uparrow}^{\dagger}c_{i\downarrow}^{\dagger}+\Delta_{i}^{*}c_{i\downarrow}c_{i\uparrow})
    \label{hammf}
\end{eqnarray}
where the superconducting OP is given by $\Delta_{i} = - U \sum_{i} \langle c_{i\downarrow}c_{i\uparrow} \rangle$. The diagonalization of the mean-field Hamiltonian can be done by the Bogoliubov transformation,
\begin{eqnarray}
c_{i\sigma} = \sum_{n}(u_{i\sigma}^{n} \gamma_{n} - \sigma v_{i\sigma}^{n*} \gamma_{n}^{\dagger} ) \\
    c_{i\sigma}^{\dagger} = \sum_{n}(u_{i\sigma}^{n*} \gamma_{n}^{\dagger} - \sigma v_{i\sigma}^{n} \gamma_{n} )
    \label{uv}
\end{eqnarray}
where the spin degree of freedom is denoted by $\sigma = \pm 1$. $\gamma_{n} (\gamma_{n}^{\dagger})$ are the Bogoliubov quasiparticle annihilation (creation) operators with $u_{i\sigma}^{n} (v_{i\sigma}^{n})$ being the electron and hole-like wavefunctions of the quasiparticles with the quantum number $n$. 
The local OP, $\Delta_{i}$, must satisfy the self-consistency condition,
\begin{equation}
    \Delta_{i} = -U \sum_{n} (u_{i\uparrow}^{n} v_{i\downarrow}^{n*} f(E_{n}) - u_{i\downarrow}^{n} v_{i\uparrow}^{n*} (1-f(E_{n}))
    \label{self-cons}
\end{equation}
where $E_{n}$s are the positive energy eigenvalues and $f(E_{n})$ is the Fermi-Dirac distribution. Our calculations are carried out using the self-consistent procedure. We begin with an initial guess value of $\Delta_{i}$ and calculate $u_{i,\sigma}^{n}$ and $v_{i,\sigma}^{n}$ by diagonalizing the mean-field Hamiltonian of Eq.~\eqref{hammf}, and then find the $\Delta_i$ self-consistently. During each iteration, we compare the absolute difference of the initial and final value of the OP at every site with the tolerance of $10^{-5}$.
We also define the average OP as 
\begin{equation}
\langle \Delta \rangle =\frac{1}{N}\sum_{i} \Delta_{i}
\label{Eq:avgOP}
\end{equation}
where $N$ is the total number of sites in each ABS.

The electronic specific heat is defined via the  thermodynamic relation,
\begin{equation}
    C_e(T) = T \frac{dS(T)}{dT}
    \label{Eq:Cev_T}
\end{equation}
where $ S(T) $  denotes the temperature-dependent entropy of the system. The entropy takes the form~\cite{Wang2024},
\begin{equation}
    S(T) = 2 k_{B} \sum_{n} \left[ \ln\left(1 + e^{-\beta E_n} \right) + \frac{\beta E_n}{e^{\beta E_n} + 1} \right]
\end{equation}
where the summation runs over the eigenvalues $E_n $ and $ \beta = 1/(k_B T)$.

For extending our analysis to the $2$D ABSs family, the tight-binding Hamiltonian accommodates additional hopping along the vertical directions, say, $t_v$ between adjacent layers, in addition to the two horizontal hoppings $t_A$ and $t_B$.

Throughout the present study, we use the natural unit where the Boltzmann constant $k_B=1$ and scale $U$ by the hopping strength, which is in the units of the energy. We do all calculations for the half-filling and set the onsite potential $\epsilon$ to zero.

\section{Results and discussions}
\label{III}

In this section, we present and discuss our numerical results for both $1$D and $2$D ABSs. Since our main results for $2$D ABSs qualitatively corroborate with those of $1$D versions, we start with the results for $1$D aperiodic binary chains for pedagogical understanding, followed by results for $2$D ABSs for the suitability of the mean-field calculation. 

\subsection{One-dimensional aperiodic binary chains}
We start with $1$D ABSs with growth rules following Table-\ref{table} as prototype examples for our analysis.

\subsubsection{ABS-I and ABS-II }
In this subsection, we focus on ABS-I and ABS-II to illustrate the behavior of the OP and thermodynamic properties. While Fibonacci chain has already been studied in the literature~\cite{Sun2024, Wang2024, Sandberg2024, Kobialka2024, Rai2019,Rai2020b}, we include them for the sake of comparison and completeness. We investigate our results for two contrasting parameter regimes: (i) $t_{A} < t_{B}$ ($t_{A} = 0.25, t_{B} =1$) and (ii) $t_{A} > t_{B}$ ($t_{A} = 1, t_{B} =0.25$) keeping $U$ ($=1$) comparable to the larger hopping.

\begin{center}
{\it 1.1$~$ Spatial profile of OP}
\end{center}
We refer to Fig.~\ref{OI1} for the distribution of the local OP ($\Delta_{i}$) in G$_{12}$ Fibonacci ($234$ sites), G$_{7}$ Silver-mean ($240$ sites), and G$_{5}$ Bronze-mean ($143$ sites) QPS\footnote{The generations of ABSs are selected to maintain consistency in chain lengths across different structures.}.\label{footnote-size} For comparison, we also show results for the periodic counterpart for two limits of the hopping parameters. In periodic systems, the OP maintains a uniform (constant) value across all sites because of the translational symmetry present in the lattice. However, in ABSs, due to the lack of transitional symmetry, the behavior of OP becomes position-dependent. The number of branches in OP along with the ratio of their magnitudes are sensitive to the growth rule and the choice of parameters, maintaining the self-similar behavior, a hallmark of quasiperiodicity.

\begin{figure}
\centering
\includegraphics[width=\linewidth]{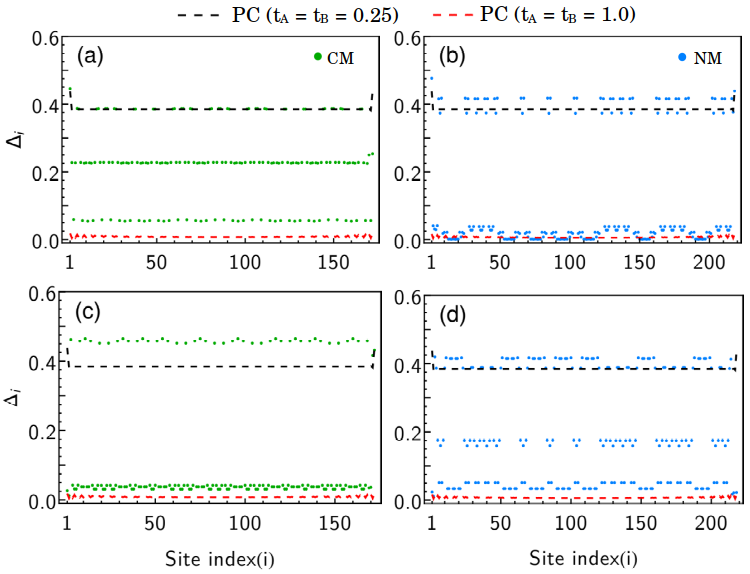}
\caption{Distribution of local OP over the sites of (a,c) Copper-mean (CM) and (b,d) Nickel-mean (NM) ABSs, which are in ABS-II family for (a-b) $t_{A} = 0.25$, $t_{B} = 1$ and (c-d) $t_{A} = 1$, $t_{B} = 0.25$. The rest of the parameter values and abbreviations are set as the same in Fig.~\ref{OI1}.}  
\label{OI2}
\end{figure}

In Fig.~\ref{OI1} (a-c), we observe that for $t_{A}<t_{B}$ the OP profiles of the ABS-I consist of two distinct branches, lower (very close to zero) and upper with some fluctuations due to structural frustration. This two-branching phenomenon of the OP is also true for the opposite parameter regime, $t_{A}>t_{B}$ (see Fig.~\ref{OI1}(d-f)) with reduced amplitude in the upper branch. Thus, within the ABS-I family, the Bronze-mean ABS exhibits the lowest OP, indicating very low pair formation, since the total number of sites participating in the formation of the upper branch is very small compared to that in the other two QPSs.
Next, we turn our attention to the ABS-II family: $G_{8}$ Copper-mean ($172$ sites) and $G_{7}$ Nickel-mean ($218$ sites) ABS. Interestingly, for  $t_A<t_B$ the Copper-mean ABS, three branches in the OP profile in contrast to the Nickel-mean chain (see Fig.~\ref{OI2}(a-b)).
The behavior of the OP in the two ABS-II chains is reversed in the opposite parameter regime ($t_A>t_B$) as shown in Fig.~\ref{OI2}(c-d). The values of OPs lie within the limits set by the periodic chains with $t_A=t_B=1$ and $t_A=t_B=0.25$ (see Fig.~\ref{OI1}).

\begin{figure}
\centering
\includegraphics[width=0.8\linewidth]{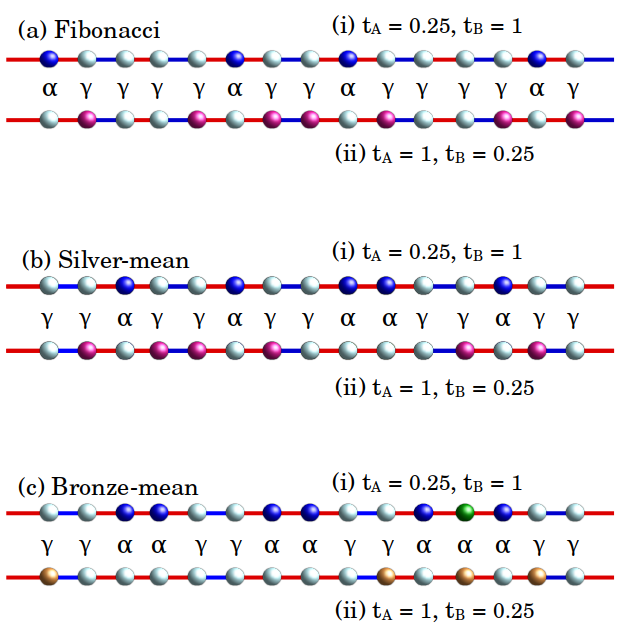}\\
\vspace{0.8cm}
\includegraphics[width=0.8\linewidth]{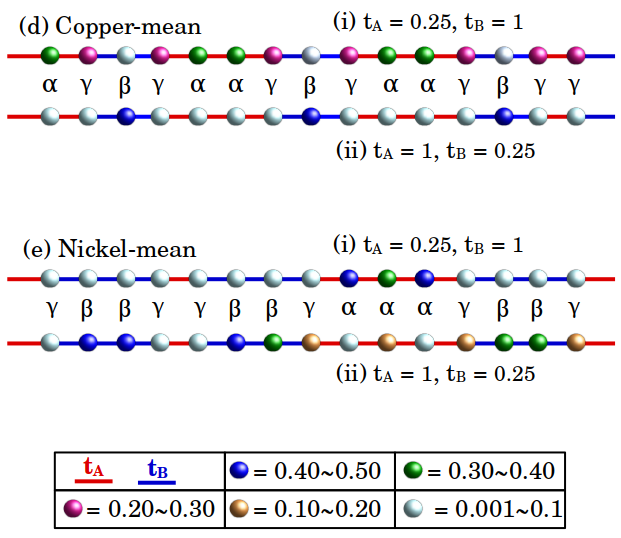}
\caption{Color-coding of various sites $\alpha$, $\beta$, and $\gamma$, according to the distribution of OP across sites of ABS-I (QPS) and ABS-II for the two parameter regimes. The rest of the parameter values are kept the same as in Fig.~\ref{OI1}.}
\label{OI_C}
\end{figure}

Now, the question is: what underlies the behavior of the OP in these ABSs? To analyze it, we identify three different kinds of sites depending on their local environment and mark them as: $\alpha$ sites flanked between $t_A$ - $t_A$, $\beta$ sites within $t_B$ - $t_B$, and $\gamma$ sites located within the environment $t_A$ - $t_B$  or $t_B$ - $t_A$ pairs as shown in Fig.~\ref{OI_C}. Remarkably, $\beta$ sites are not permissible by the recursive growth rules of ABS-I. For $t_{A} < t_{B}$, the OP amplitudes are high at all $\alpha$ sites of ABS-I, indicating the higher possibility of Cooper pair formation since the mobility of electrons in either direction is low at these sites. Importantly, the on-site attractive Hubbard interaction $U=1$ ($\sim t_B$ but $\gg t_A$) plays an important role here. Due to higher $U$ compared to the local hoppings at all $\alpha$ sites, the electron pairs are always stuck here compared to other sites. The OP distributions are marked by distinct colors in Fig.~\ref{OI_C}(a-c)(i) with some local fluctuations. In contrast, for $t_A>t_B$, OP amplitudes are lower at all $\alpha$ sites in all ABS-I (see Fig.~\ref{OI_C}(a-c)(ii)) since electrons are likely to hop to their neighboring sites instead of pair formation. At the same time, $\gamma$ sites are not at the same standard to form the pair. Specifically, $\gamma$ sites only within $\gamma$-$\alpha$-$\gamma$ environments host higher OP since for the given parameter regime, the hopping integrals between $\alpha$ and $\gamma$ sites are higher allowing pair formation. 
Following this analysis, we can explain OP behaviors across all other ABSs. Note that, in the Silver-mean ABS, an additional type of cluster, $\gamma$-$\alpha$-$\alpha$-$\gamma$, can appear, which exhibits reduced OP (see Fig.~\ref{OI_C}\,(b:ii)). In the Bronze-mean ABS, $\gamma$-$\alpha$-$\alpha$-$\gamma$ and $\gamma$-$\alpha$-$\alpha$-$\alpha$-$\gamma$ are only allowed clusters to appear; the $\gamma$-$\alpha$-$\gamma$ cluster never appears. These structural constraints lead to an overall lower OP profile in the Bronze-mean ABS. Within the $\gamma$-$\alpha$-$\alpha$-$\alpha$-$\gamma$ cluster, the middle $\alpha$ sites and the two outer $\gamma$ sites can have lower but finite amplitudes of the OP (see Fig.~\ref{OI_C}\,(c:ii)). We refer to Table-\ref{table-OP} in Appendix~\ref{Appendix_OPreal} for the summary of the OP behaviors.

Till now, our whole analysis is based on the real-space analysis in terms of $\alpha$, $\beta$, and $\gamma$ sites. It can also be explained using a conumber space description, where the sites with equivalent environments are grouped together. As an illustrative example, we compute the OP distribution in the conumber space of the Fibonacci QPS and show in Fig.~\ref{Conumber}(a-b) of Appendix~\ref{Appendix_conumber}. The behavior of the OP can be explained by the spectral weight. Due to the variations in local environments, the eigenstates are not uniformly distributed across lattice sites depending on the parameter regime (see Fig.~\ref{Conumber}(c-d)). Note that, when $t_A < t_B$, all $\alpha$ sites containing weak bonds on each side and $\gamma$ sites with one strong and one weak bond on both sides are termed as atomic and molecular sites, respectively, in the literature~\cite{Wang2024}. At atomic sites, the spectral weights are larger than those at molecular sites, which helps form pairs, as reflected in the OP distribution. In the reverse parameter regime, the higher spectral weight is not hosted by $\alpha$ sites but by $\gamma$ sites. We refer to Appendix~\ref{Appendix_conumber} for detailed discussion. An interesting point to note here, the superconducting gap is the global property of the ABS. It is fixed for all sites as seen from the local density of states (DoS) of Fibonacci QPS in Ref.~\cite{Wang2024}. 
\begin{figure}
\centering
\includegraphics[scale=0.21]{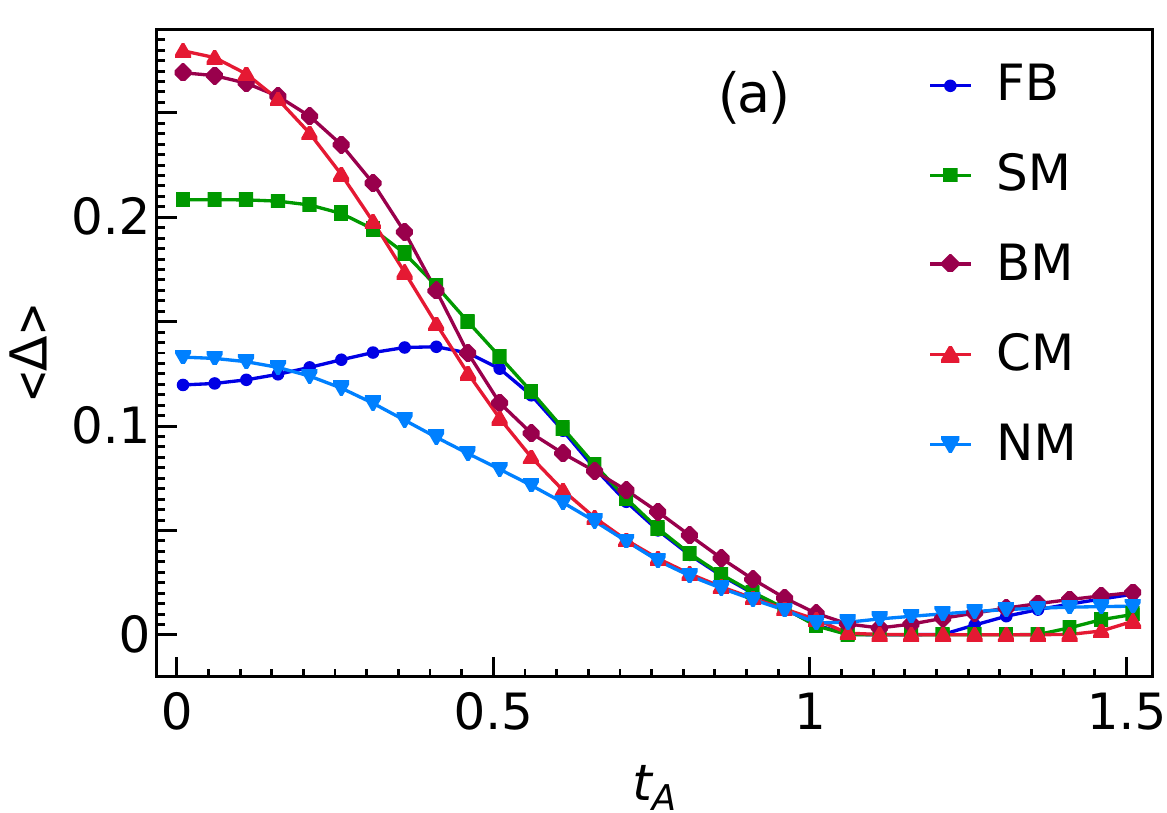}
\includegraphics[scale=0.21]{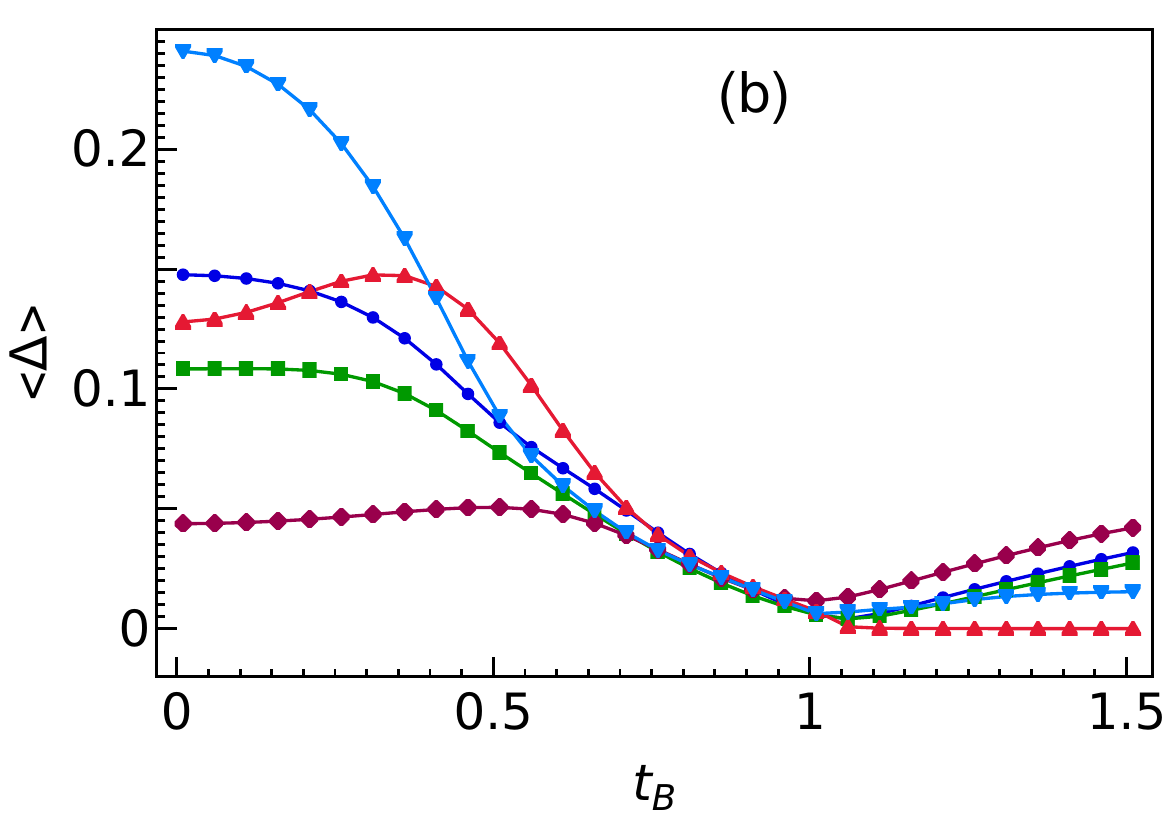}
\caption{Behavior of average OP as a function of hopping strength (a) $t_A$ keeping $t_B=1$ and (b)  $t_B$ with $t_A$ for ABS-I (QPS) and ABS-II class members. The rest of the parameter values are maintained as in Fig.~\ref{OI1}. The abbreviations are as follows: FB: Fibonacci QC, SM: Silver-mean QPS, BM: Bronze-mean QPS, CM: Copper-mean ABS, and NM: Nickel-mean ABS.}
\label{tatb}
\end{figure}

Now, we move on to the ABS-II family. Note that, all three kinds of sites, $\alpha$, $\beta$, and $\gamma$, are present in ABS-II in contrast to ABS-I. For $t_{A} < t_{B}$, relatively higher OP appears at all $\alpha$ sites, while OPs at $\beta$ sites are vanishingly small as seen Fig.~\ref{OI_C} (d-e)(i). Additionally, electrons that participate in Cooper pair formation have a higher chance to move to neighboring $\gamma$ sites of $\gamma$-$\beta$-$\gamma$ cluster in the Copper-mean ABS and thus host finite OP but still lower compared to $\alpha$ sites. Interestingly, the breaking of the finite OP in two branches undergoes a noticeable change when we consider the reverse hopping parameter regime (see Fig.~\ref{OI_C} (d-e)(ii)) and can be explained in a similar way. However, in Nickel-mean ABS, the middle $\alpha$ and outer $\gamma$ sites of $\gamma$-$\alpha$-$\alpha$-$\alpha$-$\gamma$ clusters host relatively lower OP and form a middle branch. We refer to Table-\ref{table-OP} for the summary. Note that, in the ABS-II family, for $t_A<t_B$, the $\alpha$ sites are the atomic sites and exhibit higher spectral weight, while in the reverse parameter regime, $\beta$ sites hosting higher spectral weight as well as higher OP amplitudes play the role of atomic sites. Importantly, the behaviors of OP will remain qualitatively similar for low to moderate $U$, but will break down for strong $U$. The Hartree shift term is not taken into account for all these results in the main text. The effect of the Hartree shift is discussed in Appendix~\ref{Appendix_Hartree_shift}.

Till now, behaviors of the OP have been discussed for two set of parameter values. In Fig.~\ref{tatb}(a) (Fig.~\ref{tatb}(b)), we plot the average OP as a function of $t_A$ ($t_B$), keeping the other one fixed at unity i.e., $t_B=1$ ($t_A=1$), keeping $U=1$. The ratio of the Hubbard strength to the hopping integrals plays the central role here. Comparing ABS-I and ABS-II family members in Fig.~\ref{tatb}(a), we find that the Fibonacci and Nickel-mean ABS show lower average OP, while the Bronze mean and Copper-mean ABS exhibit the higher values when $t_{A} < t_{B}=U$ as discussed previously. This is due to the higher spectral weight at the $\alpha$ sites, and Bronze-mean contains the maximum $\alpha$ sites among all ABSs. Now, on further tuning the hopping parameter, $t_A$ reaches the value of $t_B$, i.e., $t_A=t_B=1$, which describes the periodic limit, and all OP curves converge to a single point. Further increment of $t_{A}$ not only reaches the contrast parameter regime ($t_A>U>t_B$) but also makes $t_A$ higher than $U$, which eventually leads to suppressed pair formation with much lower average OP. In Fig.~\ref{tatb}(b), as we tune $t_{B}$ maintaining $t_{A}=U> t_{B}$, the Nickel-mean ABS exhibits the highest average OP as a large number of $\beta$ sites appear here. However, in the Bronze-mean ABS, the average OP is very small due to the absence of those sites.
\begin{figure}
\centering
\includegraphics[scale=0.32]{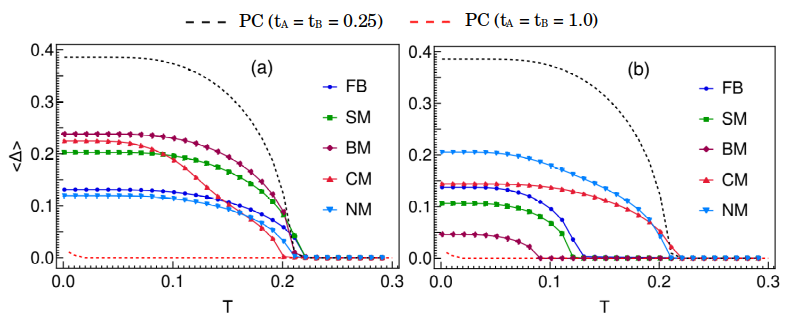}
\caption{Variation of average OP with temperature for the ABS-I and ABS-II for different choices of hopping amplitudes (a) $t_{A} = 0.25, t_{B} = 1$, (b) $t_{A} = 1, t_{B} = 0.25$. The rest of the parameter values are maintained as in Fig.~\ref{OI1} and the abbreviations are the same as in Fig.~\ref{tatb} except PC, which indicates periodic case.}
\label{OT}
\end{figure}

\begin{center}
{\it 1.2$~$ Temperature-dependence of OP}
\end{center}
Now, an interesting and important question is: how does the OP change with temperature? In periodic systems, the OP typically shows a smooth and well-understood decreasing behavior with increasing temperature and eventually vanishes at a well-defined critical temperature $T_{c}$ (see black dotted curves in Fig.~\ref{OT}). This critical point marks a phase transition beyond which the system loses its pair formation phase, and the average OP drops to zero. The lack of translational symmetry affects the critical temperature value. However, the appearance of a single critical temperature for all these ABSs indicates the bulk transition irrespective of the variation of the local OP supporting the discussions in Ref.~\cite{Wang2024}. The critical temperature depends on the superconducting gap, which varies with the parameter regime, as seen in the DoS profiles in Appendix~\ref{Appendix_Filling}. 
To illustrate further, for $t_A<t_B$, all OP curves go down to zero around similar temperatures, highlighting similar transition temperatures (see Fig.~\ref{OT}(a)). For this parameter regime, the widths of the superconducting gaps are very similar for different ABSs as shown in Fig.~\ref{mu} of Appendix~\ref{Appendix_Filling}, and thus, the corresponding transition temperatures appear at very close values. However, the OP profile for the Copper-mean ABS deviates from the usual dome-like structures. This can be understood from the individual behavior of different branches of the OP profiles. We check that the middle branch decreases rapidly compared to the upper one and also the other ABSs.
\begin{figure}
\centering
\includegraphics[width=\linewidth]{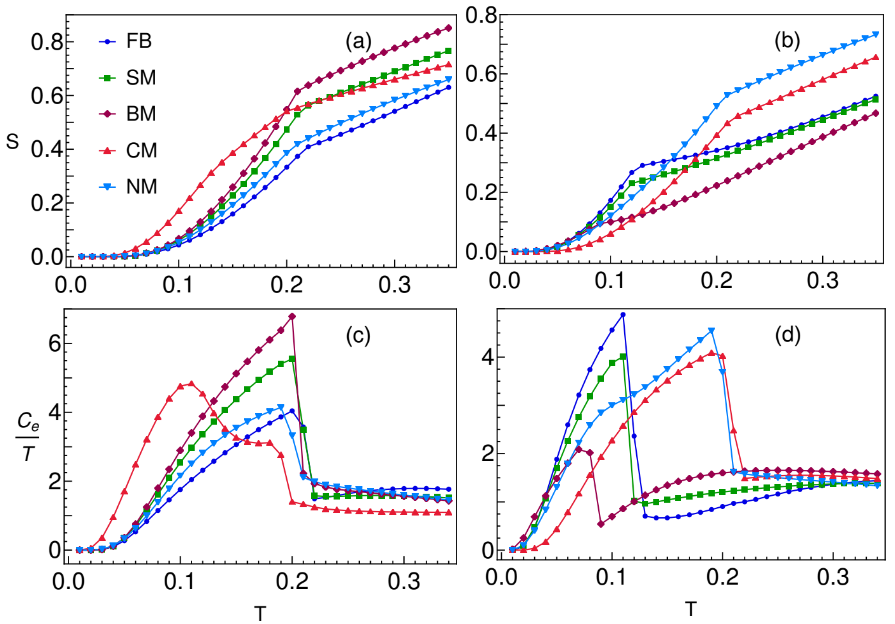}
\caption{(a,b) Entropy and (c,d) electronic specific heat vs. temperature for the ABS-I and ABS-II with (a,c) $t_{A} = 0.25$, $t_{B} = 1$ and (b,d) $t_{A} = 1$, $t_{B} = 0.25$. The rest of the parameter values are maintained as in Fig.~\ref{OI1} and the abbreviations are the same as in Fig.~\ref{tatb}.}
\label{En_Ce}
\end{figure}
For the reverse situation ($t_{A} > t_{B}$), the ratios of the OP amplitudes found in Fig.~\ref{OT}(b) follow the discussions in the previous subsection. Importantly, the transition temperatures for various ABSs are now different from each other. This can be explained by the superconducting gap widths shown in the DoS profiles in Fig.~\ref{mu} of Appendix~\ref{Appendix_Filling}. Systems containing larger superconducting gap widths exhibit higher transition temperatures and vice versa.

On the whole, the variation of the average OP with the temperature is not universal but shows distinct behaviors depending on the underlying growth rule and the specific choice of model parameters. Comparison of the OP profiles and the transition temperatures across various ABSs identifies which configuration strongly supports Cooper pair formation. This understanding is not only fundamental for understanding superconducting phases in such aperiodic systems, but also lays the groundwork for future research aimed at a better understanding of the ABS-based superconductivity.

\begin{center}
{\it 1.3$~$ Thermodynamic properties}
\end{center}
We now focus on the thermodynamic behavior of these ABSs in terms of the entropy and electronic specific heat, which serves as a sensitive probe of the quasiparticle excitation spectrum and provides information about the possible superconducting phases. The behaviors of entropy $S(T)$ and electronic specific heat $C_{e} (T)/T$ with temperature are shown in Fig.~\ref{En_Ce} for ABS-I and ABS-II family members,  considering the two parameter regimes. At low temperatures, the state is highly ordered, with most electrons forming Cooper pairs, resulting in lower entropy. As temperature increases, thermal excitations break pairs resulting in increasing disorder and entropy. The electronic specific heat profiles drop sharply at the transition temperatures for all of the ABSs. This sudden change is a clear sign of the normal to superconducting phase transition. Above the transition temperature, the specific heat behaves more smoothly. This is the typical behavior found in ordinary Bardeen-Cooper-Schrieffer (BCS) superconductors. We avoid repeating the BCS behaviors in more detail, rather explain only the deviations from the BCS behaviors due to the aperiodicity.
\begin{figure} 
\centering
\includegraphics[width=\linewidth]{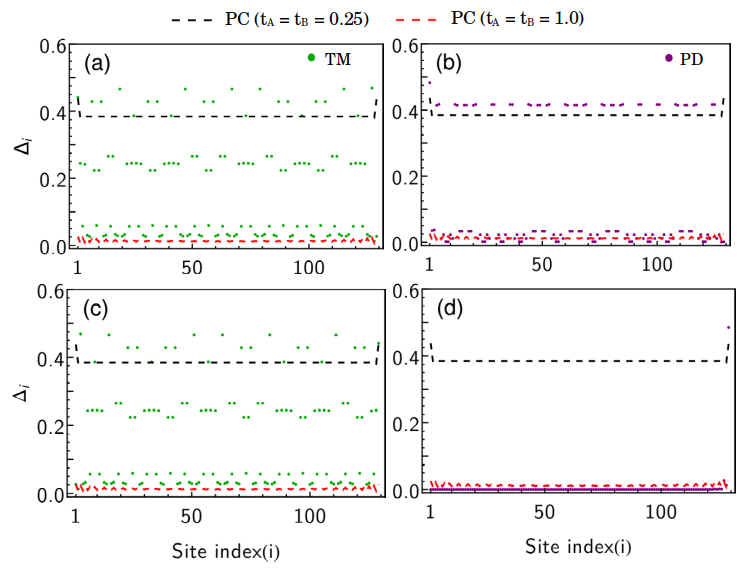}
\caption{Local OP amplitudes at various sites in $1$D (a,c) Thue-Morse (TM) and (b,d) Period-doubling (PD) ABSs, which are in ABS-III family. The hopping terms are chosen as: (a,b) $t_{A} = 0.25, t_{B} = 1$, (c,d) $t_{A} = 1, t_{B} = 0.25$. The rest of the parameter values and abbreviations are set as the same in Fig.~\ref{OI1}.}
\label{OI-TM-PD}
\end{figure}

However, there are some changes found in the behavior of both the entropy and specific heat when we switch the parameter regimes. Specifically, for $t_A<t_B$, all members of the ABS-I and ABS-II family show a similar pattern for the temperature dependence except the Copper-mean ABS (see Fig.~\ref{En_Ce}(a,c)). The reason behind this unique thermodynamic behavior of the Copper-mean ABS is hidden in its OP distribution, or equivalently, the structural growth. As discussed earlier, the non-zero OP profile of the Copper-mean ABS breaks into two parts: the higher one is for $\alpha$ sites, and the other is for the $\gamma$ sites of the $\gamma$-$\beta$-$\gamma$ cluster (see Fig.~\ref{OI2} (a)). Now, as the temperature rises, the OP magnitudes of both branches decrease, and the lower branch goes to zero, whereas the higher branch still contains non-zero values. It approaches zero with further increasing temperature. Due to this, the entropy and electronic specific heat for the Copper-mean ABS vary in an unusual way. The first decay of $C_{e}$ from the peak is due to the suppression of OP values at $\gamma$ sites of $\gamma$-$\beta$-$\gamma$ cluster, the second and final fall indicates the disappearance of the OP at $\alpha$ (and also from $\gamma$) sites (see Fig.~\ref{En_Ce}(c)). Following the same reason, the rapid fall of the average OP with temperature for Copper-mean ABS (see Fig.~\ref{OT}(a)) can be explained.

For the contrast parameter regime, the behaviors of the entropy and the specific heat for the ABS-I and ABS-II family are shown in Fig.~\ref{En_Ce}(b,d). Interestingly, the profiles show phase transition at different $T_c$, which corroborate our findings mentioned in the previous subsection. Note that, the behavior of the specific heat profile for the Nickel-mean ABS shows different behavior from that of the others. It is characterized by a small curvature, which can be explained in the following way. Among the two non-zero OP branches of the Nickel-mean ABS, the lower branch goes down to zero rapidly, followed by the decreasing behavior of the other branch. This two-step process is reflected in the curvature profile (see Fig.~\ref{En_Ce}(d)). For the analysis of the deviation from the periodic limit, we show a comparison with the periodic system in Appendix~\ref{Appendix_periodic_limit}. We also compare these thermodynamic quantities calculated for $U = 0$ and $U \ne 0$ for a better understanding.
\begin{figure}
\centering
\includegraphics[width=0.8\linewidth]{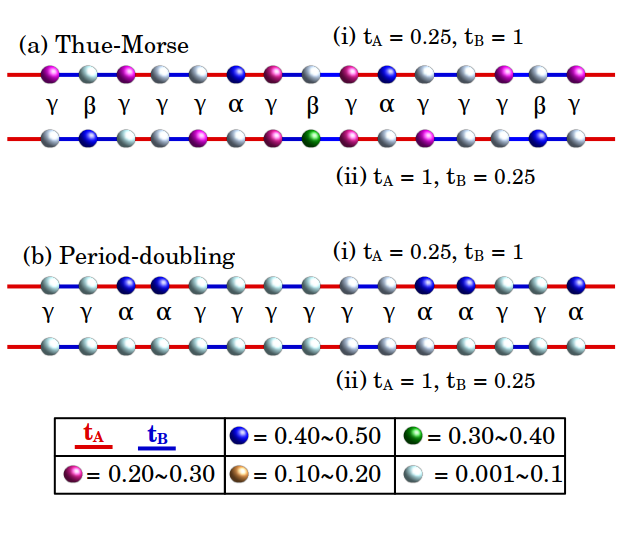}
\caption{Color-coded distribution of OP amplitudes in (a) Thue-Morse and (b) Period-doubling ABSs, which are in ABS-III family. The rest of the parameter values are set as the same in Fig.~\ref{OI1}.}
\label{OI-fig-TM-PD}
\end{figure}
\begin{figure*}
\centering
\includegraphics[width=0.8\linewidth]{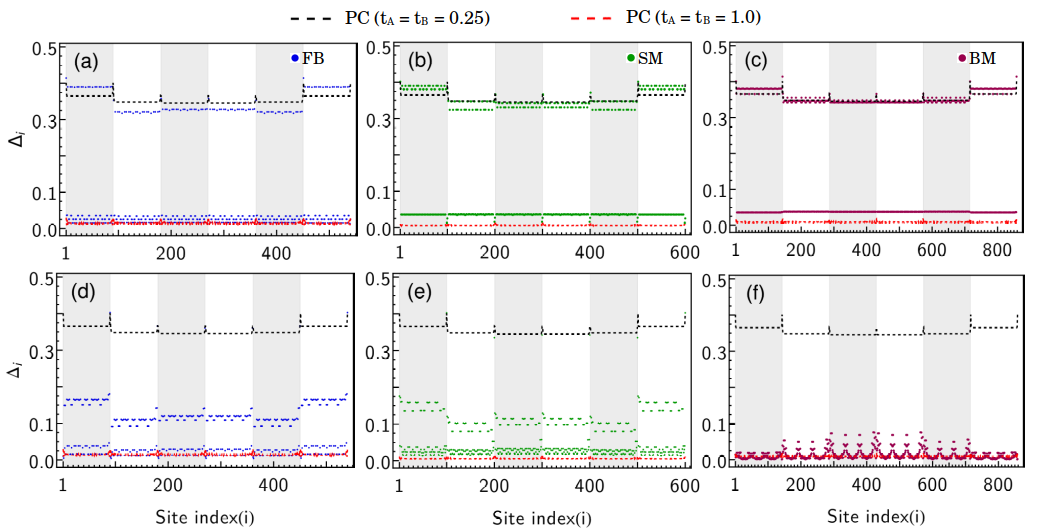}
\caption{Distribution of OP over sites in $2$D with $M\times N$ sites (a,d) Fibonacci ($M=6,N=90$), (b,e) Silver-mean ($M=6,N=100$), (c,f) Bronze-mean ($M=6,N=143$) QPS for $t_v=0.25$ and (a-c) $t_A =  0.25$, $t_B = 1$ and (d-f) $t_A=1$, $t_B=0.25$. OP profiles for two periodic limits are shown by black ($t_A=t_B=0.25$) and red ($t_A=t_B=1$) dotted lines. Other parameters and abbreviations are the same as in Fig.~\ref{OI1}. We mark odd M ($M =1, 3, 5$) by gray shades.} 
\label{OI_2D_FIBONACCI1}
\end{figure*}

\subsubsection{ABS-III}
For the completeness of our work, we now extend our investigation beyond the ABS-I (QPS) and ABS-II, with distinct geometric and spectral features (see Table~\ref{table}), to show that our analysis is also valid for other types of ABSs. To minimize repetitions of analysis, we only choose two ABSs: Thue-Morse and Period-doubling sequence which we call as ABS-III, and explain only the behaviors of the OP in the main text. We consider both these ABS in their $G_8$-th generation, containing $129$ sites.

\begin{center}
{\it 2.1$~$ Behavior of OP}
\end{center}
Figure~\ref{OI-TM-PD} shows the variation of the local OP over sites for the Thue-Morse and period-doubling ABS. In the Thue-Morse ABS, all three $\alpha$, $\beta$, and $\gamma$ site appear similar to the ABS-II family. When $t_A<t_B$, all $\alpha$ sites exhibit higher OP values (see Fig.~\ref{OI-fig-TM-PD} (a:i)). On the other hand, the OP amplitudes at $\beta$ sites are suppressed due to the higher hopping probability to their neighbor $\gamma$ sites. Hence, these $\gamma$ sites of $\gamma$-$\beta$-$\gamma$ cluster exhibit a finite OP (low compared to $\alpha$ sites). The OP profiles for these ABSs are depicted in Fig.~\ref{OI-TM-PD}(a), showing two finite OP branches for $\alpha$ sites and for $\gamma$ sites within the $\gamma$-$\beta$-$\gamma$ clusters.

For the reverse situation $t_A>t_B$, the distribution of the OP is completely reversed. Now $\alpha$ sites contain the lowest OP amplitude, whereas the OP amplitudes at the $\beta$ sites are higher. Additionally, the $\gamma$ sites exhibit a finite OP (lower compared to $\beta$ sites), which comes from $\gamma$-$\alpha$-$\gamma$ cluster, not from $\gamma$-$\beta$-$\gamma$ cluster (see Fig.~\ref{OI-TM-PD}(c)). We present this by color-coding in Fig.~\ref{OI-fig-TM-PD}(a:ii). 

For the period-doubling ABS, the formation of the $\beta$ site is never possible as found in the ABS-I family. At $t_{A}<t_{B}$, only $\alpha$ sites have the higher OP amplitudes as shown in Fig.~\ref{OI-TM-PD}(b) and Fig.~\ref{OI-fig-TM-PD}(b:i). When the period-doubling ABS is subjected to the reverse parameter regime, i.e., $t_{A}>t_{B}$, the OP at all sites becomes very close to zero. Due to the formation of the $\beta$ sites, as well as $\gamma$-$\alpha$-$\gamma$ (present in the Fibonacci and Silver-mean QPS too) or $\gamma$-$\alpha$-$\alpha$-$\alpha$-$\gamma$ (present in Bronze-mean and Nickel-mean ABS too) cluster is never possible here. The OP profile and the corresponding schematic diagram for the period-doubling ABS are depicted in Fig.~\ref{OI-TM-PD}(d) and Fig.~\ref{OI-fig-TM-PD}(b:ii), respectively. We denote the site types and show the OP amplitudes comparison in Table-\ref{table-OP2}(see Appendix~\ref{Appendix_OPreal}). We find that the behavior of the average OP with temperature and thermodynamic behaviors for these members of ABS-III family are close to the ABS-I and ABS-II, which we explicitly present in Appendix~\ref{Appendix_ABSIII}.

\subsection{Two-dimensional aperiodic binary systems}
\begin{figure}
\centering
\centering
    \includegraphics[width=\linewidth]{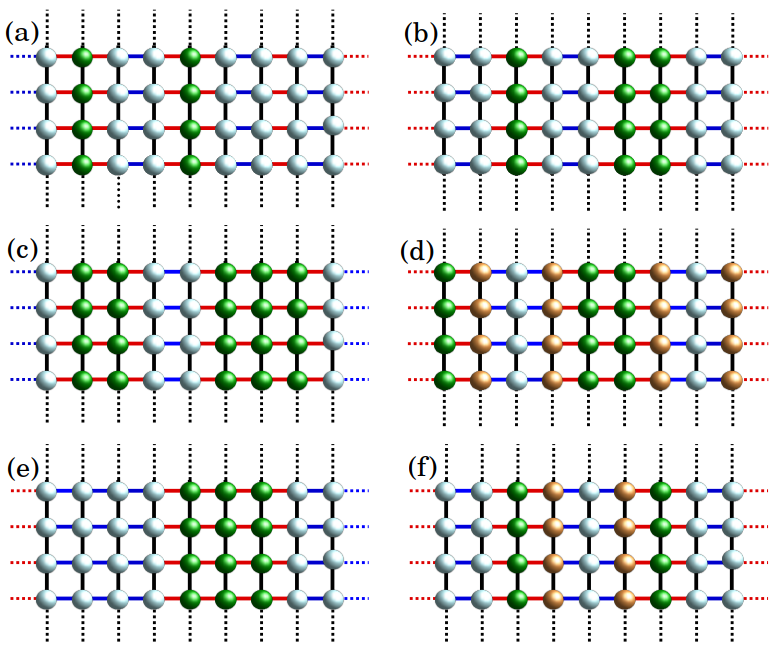}
    \includegraphics[width=\linewidth]{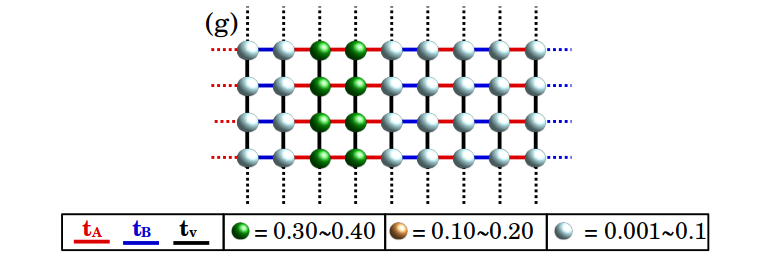}
\caption{Behavior of OP in $2$D (a) Fibonacci, (b) Silver-mean,  (c) Bronze-mean, (d) Copper-mean, (e) Nickel-mean, (f) Thue-Morse, and (g) Period-doubling ABSs with the local OP distribution marked by using color code. The horizontal hopping is $t_{A}$ or $t_{B}$ depending on the growth rule, whereas along the vertical, the hopping term is $t_{v}$. The parameters are chosen as $t_{A} = 0.25, t_{B} =1$, $t_{v} =0.25$. The rest of the parameter values are set as the same in Fig.~\ref{OI1}.  }
\label{2DQC}
\end{figure}

Finally, we extend our investigation to $2$D ABSs, keeping in mind the validation of the mean-field analysis, and present our results for the OP in Fig.~\ref{OI_2D_FIBONACCI1}. We construct the $2$D lattices by vertically stacking $M$ number of $1$D chains as shown in Fig.~\ref{2DQC}. Unlike $1$D cases, where only two types of overlap integrals appear for all growth rules we consider, the $2$D structure allows increased complexity in the local bonding environments due to the additional hopping. Specifically, three distinct types of overlap integrals can appear, denoted by $t_A$, $t_B$, and $t_v$, each of them corresponding to different geometric arrangements of neighboring sites. Interestingly, our analysis in the $2$D setting not only aligns with but also reinforces our key findings obtained from the $1$D case, highlighting the robustness of the observed physical behaviors of OP distributions across different dimensions. We consider $2$D $G_{10}$-th Fibonacci QPS containing $540$ sites, $G_6$-th generation Silver-mean QPS with $600$ sites, and $G_5$-th generation Bronze-mean QPS consisting of $858$ sites for the two different parameter regimes. 

Figure~\ref{OI_2D_FIBONACCI1} illustrates the spatial distribution of the OP across the $2$D lattice. 
Along the $x$-axis, we consider $M\times N$ total number of sites where $N$ is the size of the generation.\footnote{For example, in $2$D Fibonacci QPS, we take total $6$ number of $1$D Fibonacci chains of $G_{10}$ generation (each chain containing $90$ sites). The OP for $2$D Fibonacci QPS (see Fig.~\ref{OI_2D_FIBONACCI1}(a,d)) is plotted against the site index of the entire lattice, where site index $1$ to $90$ indicates the site index of the first chain, $91$ to $180$ for the second chain, and so on. Similar numbering is followed for other $2$D ABSs too.} We believe that these plots would appear smoother for a larger system size (by increment of both the generation index of $1$D chain and stack more such chains). However, performing such Calculations become numerically demanding and difficult to handle in practice. In Fig.~\ref{OI_2D_FIBONACCI1}, we observe that the OP profiles are more enriched than the $1$D counterpart. The OP distributions of $2$D ABS-II and III families are elaborately discussed in Appendix~\ref{Appendix_2D}. 

For $t_A<t_B$, the superconducting OP profiles of the $2$D ABS-I (QPS) members exhibit two primary branches similar to those found $1$D systems, see Fig.~\ref{OI_2D_FIBONACCI1}(a-c). Conversely, at the reverse parameter regime, i.e., $t_A>t_B$, the overall OP of the ABS-I appears at smaller amplitudes compared to the previous regime. These OP profiles are depicted in Fig.~\ref{OI_2D_FIBONACCI1}(d-f). While the overall qualitative behavior is preserved, quantitative differences arise in the OP amplitudes. This happens mostly due to the differences in the generation number and definitely the system size between the $1$D and $2$D ABSs. Due to computational resource limitations, we consider lower-generation $2$D systems compared to the $1$D counterpart, resulting in a smaller number of lattice sites along the horizontal direction, although the total number of sites in the $2$D systems is large. It is worth noting that edge effects are present in all OP profiles. Such effects have been previously discussed in the literature~\cite{Tanaka2000, Croitoru2000, Stojkoviifmmode1993, Troy1995, Martin1998, Shanenko2006,Samoilenka2020,Samoilenka2021, Hainzl2022}. We confirm that the influence of edge effects decreases as the generation number increases, leading to more uniform bulk behavior in higher-generation lattices.

However, the basic patterns of the OP profiles in the $2$D systems are very similar to those in $1$D. The corresponding schematic diagram of $2$D structures at the parameter regime $t_A<t_B$ is shown in Fig.~\ref {2DQC} for all ABS. We consider the vertical hopping term $t_v$ equal to the lower hopping of the binary systems. As discussed in $1$D case, here also the $\alpha$ sites contain the highest magnitude of OP. The corresponding sites are marked by green color as shown in Fig.~\ref{2DQC}(a-g) for $2$D Fibonacci, Silver-mean, Bronze-mean QPSs, and Copper-mean, Nickel-mean, Thue-Morse, and Period-doubling ABSs respectively. Additionally, as seen in $1$D case, here also the $\gamma$ sites in $\gamma-\beta-\gamma$ exhibit finite OP amplitudes. It is also true for $2$D Copper-mean and Thue-Morse ABSs as demonstrated in Fig.~\ref{2DQC}(d) and (f), by yellow color.
We also compare our results with the corresponding $2$D periodic lattice considering both the higher (black dots) and the lower (red dots) hoppings setting two extreme limits for the OP amplitudes. We observe that the deviations of the higher branch of the OP for the Fibonacci, Silver-mean and Bronze-mean are much smaller than that of the upper periodic limit. However, the scenario changes with the change in the parameter regime. It is similar to what we find in the $1$D case.
A natural question appears at this point: does the vertical hopping $t_v$ affect the OP distribution across the ABSs? The answer is yes, it does. Here, we show all results for $t_v$ to be equal to whichever is lower out of $t_A$ or $t_{B}$. However, when $t_v$ is very small compared to $t_A$ or $t_{B}$, electron hoppings along the vertical direction will become more restricted.  This reduced mobility enhances electron localization, which in turn facilitates the formation of Cooper pairs. As a result, the superconducting OP increases in magnitude. We also discuss the effect of the vertical hopping on the OP in Appendix~\ref{Appendix_2D}. On the other hand, if $t_v$ is increased beyond $t_A$ or $t_{B}$, electrons can move more freely in the vertical direction. This higher mobility tends to spread out the electrons, making it difficult for the pair formation. As a result, the superconducting OP decreases. In the extreme case where $t_v$ competes with $t_{A(B)}$ and $U$, the OP is suppressed across the entire ABSs and can become nearly zero at all sites, indicating that the pair formation phase is lost. This consistency across the dimensions suggests the universality of the underlying physical behavior to some extent, regardless of the dimensional extension of the ABS structures. The behavior of the average OPs with the temperature thermodynamic properties for the $2$D ABSs are demonstrated in Appendix~\ref{Appendix_2D}. We avoid further illustrating our results here since the behaviors of the curves for all ABSs are almost the same as discussed for $1$D cases, only with a marginal difference in the magnitudes.

\section{Summary and Conclusions}
\label{IV}
In summary, we have carried out a comprehensive investigation on the emergence and behavior of the superconducting OP in a broad class of $1$D ABSs, and further extended our analysis to $2$D systems. We have focused on the spatial structure of the OP, its sensitivity to system parameters, and their thermal responses. Using a self-consistency framework, we have investigated how the superconducting OP behaves across ABSs, identifying which sites support enhanced OP and which sites suppress it. This spatially resolved analysis has revealed that pair formation in ABSs is strongly influenced by the local structural motifs that arise from their underlying aperiodic order. Our analysis has demonstrated that both the nearest-neighbor hopping amplitude and the strength of the attractive on-site Hubbard interaction are crucial for the behavior of the OP. The competition between different kinds of sites or clusters underlying aperiodic geometry gives rise to a rich and nontrivial OP landscape that deviates significantly from the conventional periodic lattices. 

We have also explored the temperature dependence of the OP in ABSs for various parameter regimes. We have found that the transition temperatures vary significantly across the collection of ABSs and thus are not universal. It was found to vary significantly depending on both the model parameters and the specific aperiodic sequence. By systematically comparing the parameter regimes, we identify optimal combinations of hopping and interaction strengths that are favorable for achieving higher OP across the ABS family. Conversely, we have also pointed out the parameter regimes where the OP is significantly weakened. 

Finally, we have examined the thermodynamic properties of these ABSs in terms of the entropy and electronic specific heat, which show distinct variations depending on the specific choice of the sequence and the associated model parameters. The behaviors of these thermodynamic quantities in ABSs are consistent with the behaviors of the OP, further reinforcing our main findings.

From the experimental perspectives, these ABSs have been shown to be synthesized by the electron or single-beam holography~\cite{Vardeny2013, Matarazzo2011, Zito2008}. Various QPSs and their approximants have also been experimentally realized by proper choice of compound materials, like Al-Mg-Zn, Al-Cu-Li, Mg-Ga-Al-Zn alloys~\cite{Takeuchi1995, Windisch1994, Mizutani1993, Saito2020}. The superconducting phases in the Fibonacci QPS and its approximants have been uncovered in Al-Zn-Mg alloy~\cite{kamiya2018}. This provides a way for the experimental realizations in a broader range of aperiodic systems for both $1$D and $2$D systems including Penrose and Moire patterns in twisted materials~\cite{Li2024, Wang1987, 
Kumar1986, Ezzi2024, Gierer1998, Smerdon2008, Iwasaki2021, Hayashida2007}. 

Our findings not only enrich the theoretical understanding of the pair formation in a broad range of ABSs, but also offer valuable guidance for the experimental realization since this approach will help in determining the most promising structure for realizing the pair formation phase. These results may also serve as a useful theoretical foundation for interpreting calorimetric measurements in engineered ABSs. The appearance of multiple gaps and filling dependence can be related to the behavior of the OP which we keep as the future outlook. Note that, in the present study, we do not confirm any superconductivity. We only confirm the possibility of pair formation. It needs further investigation of the phase coherence or superfluid density, which we plan to address in a future study.

\section{Acknowledgement}
We acknowledge the Department of Space, Government of India, for all support at Physical Research Laboratory (PRL) and also the ParamVikram-1000 HPC facility at PRL, Ahmedabad, for carrying out a major part of the calculations. S.\,B.\, thanks Kuntal Bhattacharyya for some helpful discussions. P.\,D. thanks Annica M. Black-Schaffer, B.\,Andrei Bernevig, and Adolfo G. Grushin for constructive comments and suggestions, and Debmalya Chakraborty for some helpful discussions. We thank Miguel \'{A}ngel S\'{a}nchez-Mart\'{i}nez for careful reading of the manuscript and valuable comments. We also thank anonymous reviewers for their constructive comments and suggestions.

\appendix

\section{Order parameter in real space}
\label{Appendix_OPreal}
In the main text, the behavior of the OP has been portrayed in the real space. It turns out that any particular type of sites for a particular ABS may or may not host similar OP amplitudes. It depends on the sequence as well as the hopping ratio. We here provide a summary of the color coded OP distributions in Table~\ref{table-OP} for ABS-I and II, and Table~\ref{table-OP2} for ABS-III.
\begin{table}
\begin{center}
\caption{Summary of OP amplitudes in ABS-I and ABS-II with colors same as shown in Fig.~\ref{OI_C}.}
\renewcommand{\arraystretch}{1.5}
\setlength{\tabcolsep}{4pt}
\begin{tabular}{|c|c|c|c|}
\hline
\makecell{Parameter\\ regime} & ABSs & \makecell{higher OP \\ (color)} & \makecell{lower OP \\(color)}\\
\hline \hline
ABS-I& Fibonacci &  \makecell{all $\alpha$ \\ (blue)}  & \makecell{all $\gamma$\\ (cyan)}   \\
\cline {2-4}
\makecell{$t_{A}=0.25$\\ $t_{B} = 1$~~~~~}& Silver-mean & \makecell{all $\alpha$\\ (blue)} & \makecell{all $\gamma$ \\ (cyan)} \\
\cline{2-4}
& Bronze-mean &  \makecell{all $\alpha$\\ (blue)}  & \makecell{all $\gamma$\\ (cyan)} \\
\hline
ABS-I & Fibonacci  & \makecell{$\gamma$\\ (magenta) \\in $\gamma$-$\alpha$-$\gamma$}  & \makecell{all $\alpha$ \\(cyan)\\ other $\gamma$ \\(cyan)} \\
\cline {2-4}
\makecell{$t_{A}=1$~~~~~\\ $t_{B} = 0.25$}& Silver-mean & \makecell{$\gamma$ \\(magenta)\\ in $\gamma$-$\alpha$-$\gamma$} & \makecell{all $\alpha$\\ (cyan)\\ other $\gamma$ \\(cyan)} \\
\cline{2-4}
& Bronze-mean  & \makecell{$\gamma$\\ (yellow)\\ and  middle $\alpha$\\(yellow)\\ in $\gamma$-$\alpha$-$\alpha$-$\alpha$-$\gamma$} & \makecell{other $\alpha$\\ (cyan) \\ other $\gamma$ \\(cyan)} \\
\hline \hline
ABS-II & Copper-mean & \makecell{ all $\alpha$\\ (green) \\ $\gamma$ \\(magenta)\\ in $\gamma$-$\beta$-$\gamma$} & \makecell{all $\beta$\\ (cyan)\\ other $\gamma$\\ (cyan)} \\
\cline{2-4}
\makecell{$t_{A}=0.25$\\ $t_{B} = 1$~~~~~}& Nickel-mean & \makecell{all $\alpha$\\ (blue)} & \makecell{all $\beta$\\ (cyan)\\ all $\gamma$\\ (cyan)}  \\
\hline
ABS-II & Copper-mean & \makecell{all $\beta$\\ (blue)} & \makecell{all $\alpha$\\ (cyan)\\ all $\gamma$\\ (cyan)} \\
\cline{2-4}
\makecell{$t_{A}=1$~~~~~\\ $t_{B} = 0.25$}& Nickel-mean & \makecell{all $\beta$\\ (blue)\\ $\gamma$\\ (yellow)\\ and middle $\alpha$\\ (yellow) \\ in $\gamma$-$\alpha$-$\alpha$-$\alpha$-$\gamma$} & \makecell{other $\alpha$\\ (cyan)\\ other $\gamma$\\ (cyan)}\\
\hline
\end{tabular}
\label{table-OP}
\end{center}
\end{table}

\begin{table}
\begin{center}
\caption{OP distribution in ABS-III with colors same as in Fig.~\ref{OI-fig-TM-PD}.}
\renewcommand{\arraystretch}{1.5}
\setlength{\tabcolsep}{4pt}
\begin{tabular}{|c|c|c|c|}
\hline
\makecell{Parameter\\ regime} &  ABSs & \makecell{higher OP\\(color)} & \makecell{lower OP\\ (color)} \\
\hline\hline
\makecell{$t_{A}=0.25$,\\ $t_{B} = 1$} & Thue-Morse & \makecell{ all $\alpha$\\ (blue) \\ $\gamma$ (magenta)\\ in $\gamma$-$\beta$-$\gamma$} & \makecell{all $\beta$ \\(cyan) \\ other $\gamma$\\ (cyan)} \\
\cline{2-4}
& Period-doubling & \makecell{all $\alpha$\\ (blue)} &  \makecell{all $\gamma$\\ (cyan)} \\
\hline
\makecell{$t_{A}=1$,\\ $t_{B} = 0.25$} & Thue-Morse & \makecell{ all $\beta$\\ (blue) \\ $\gamma$ \\(magenta)\\  in $\gamma$-$\alpha$-$\gamma$} & \makecell{all $\alpha$\\ (cyan) \\ other $\gamma$ \\(cyan)} \\
\cline{2-4}
& Period-doubling & ----- & \makecell{all $\alpha$ \\(cyan)\\ all $\gamma$ \\(cyan)}  \\
\hline
\end{tabular}
\label{table-OP2}
\end{center}
\end{table}

\section{Order parameter in conumber space}
\label{Appendix_conumber}

In this section, for the comparison purposes, we show the OP in the conumber space which is an existing concept in the literature. For the illustration, we consider only $1$D Fibonacci QPS. The conumber indexing is obtained through the cut-and-project method. For $n$-{th} generation Fibonacci QPS with a particular site index $i$, the conumber can be defined as~\cite{Wang2024},
\begin{equation}
    c(i) = 1+i G_{n-1} \text{mod}  G_{n}
\end{equation}
where $G_{n}$ is $n$-{th} Fibonacci number and $c(i) \in [1, G_{n}]$. In the conumber space, \(\alpha\) sites located between two \(t_A\) bonds in the real space appear approximately in the range \( G_{n-2} + 1 \) to \( G_{n-1} + 1 \), whereas the remaining part mainly corresponds to the \(\gamma\) sites. 

\begin{figure}[ht!]
\centering
\includegraphics[width=\linewidth]{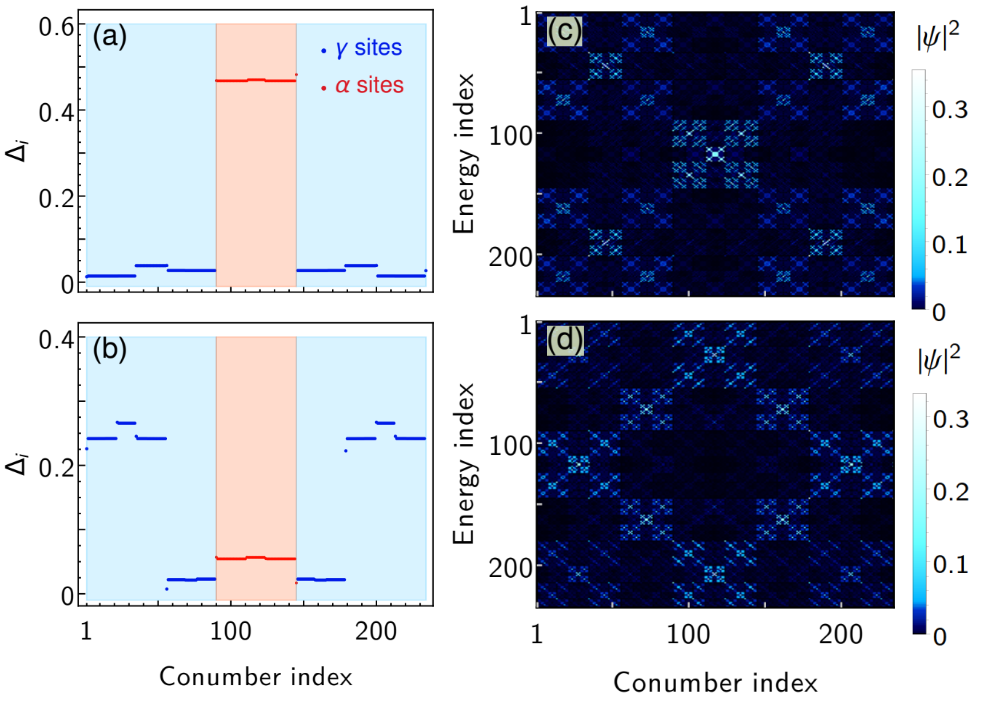}
\caption{Distribution of (a-b) OP and (c-d) spectral distributions in the conumber space for a Fibonacci chain considering (a,c) $t_{A} = 0.25$, $t_{B} = 1$ and (b,d) $t_{A} = 1$, $t_{B} = 0.25$. The rest of the parameter values are maintained as in Fig.~\ref{OI1}.}
\label{Conumber}
\end{figure}

In Fig.~\ref{Conumber}, we present the OP landscape plotted over the conumber space for the Fibonacci QPS considering two parameter regimes $(i)$ $t_A<t_B$ in Fig.~\ref{Conumber}(a) and $(ii)$ $t_A>t_B$ in Fig.~\ref{Conumber}(b). All \(\alpha\) sites are grouped around the center, and the \(\gamma\) sites occupy both sides according to their OP values. Here, we consider a $G_{12}$ Fibonacci chain containing $234$ sites. This means $\alpha$ sites are situated from $90$ to $145$ in the conumber space (marked by red in Fig.~\ref{Conumber}) and the other index corresponding to $\gamma$ sites (marked by blue in Fig.~\ref{Conumber}). We now describe the behaviors as follows. From Fig.~\ref{Conumber}(a), we can clearly see that the OP is larger at $\alpha$ sites, while the $\gamma$ sites have much smaller OPs for this parameter regime. This finding corroborates with our analysis in the main text.
 
For the reverse parameter regime, (see Fig.~\ref{Conumber}(b)), it is evident that the OP at the central region corresponding to \(\alpha\) sites (marked in red) are significantly reduced. This central part is surrounded on both sides by contributions from \(\gamma\) sites marked in blue. However, the OP does not attain higher values at all \(\gamma\) sites. Instead, the higher OP values are shown by those \(\gamma\) sites which belong to \(\gamma\)\(-\)\(\alpha\)\(-\)\(\gamma\) clusters and appear only near the two edges in the conumber space. Those $\gamma$ sites are identified from the real-space picture in the main text. There are other \(\gamma\) sites that do not belong to these specific clusters and host relatively smaller OPs. In contrast, the other \(\gamma\) sites adjacent to the central \(\alpha\)-site region show comparatively smaller OP values.

For the understanding of the behavior of the wave functions, we show spectral distributions in the conumber space in Fig.~\ref{Conumber}(c-d). For $t_A<t_B$, a higher spectral weight at all $\alpha$ (or equivalently, atomic) sites at the half filling supports the higher OP at these sites. However, in the reverse parameter regime, $\gamma$ sites exhibit higher spectral weight (see Fig.~\ref{Conumber}(d). The self-similar spectral distribution directly supports the discussions on Fig.~\ref{Conumber}(a-b) and also the discussions in the main text.
Before we leave this section, we emphasize that the results in the conumber basis fully agree with the real space description. In the conumber space, we can identify two different types of sites and arrange OP values accordingly.
\section{Effect of Hartree shift}
\label{Appendix_Hartree_shift}
\begin{figure}
\centering
\includegraphics[width=0.7\linewidth]{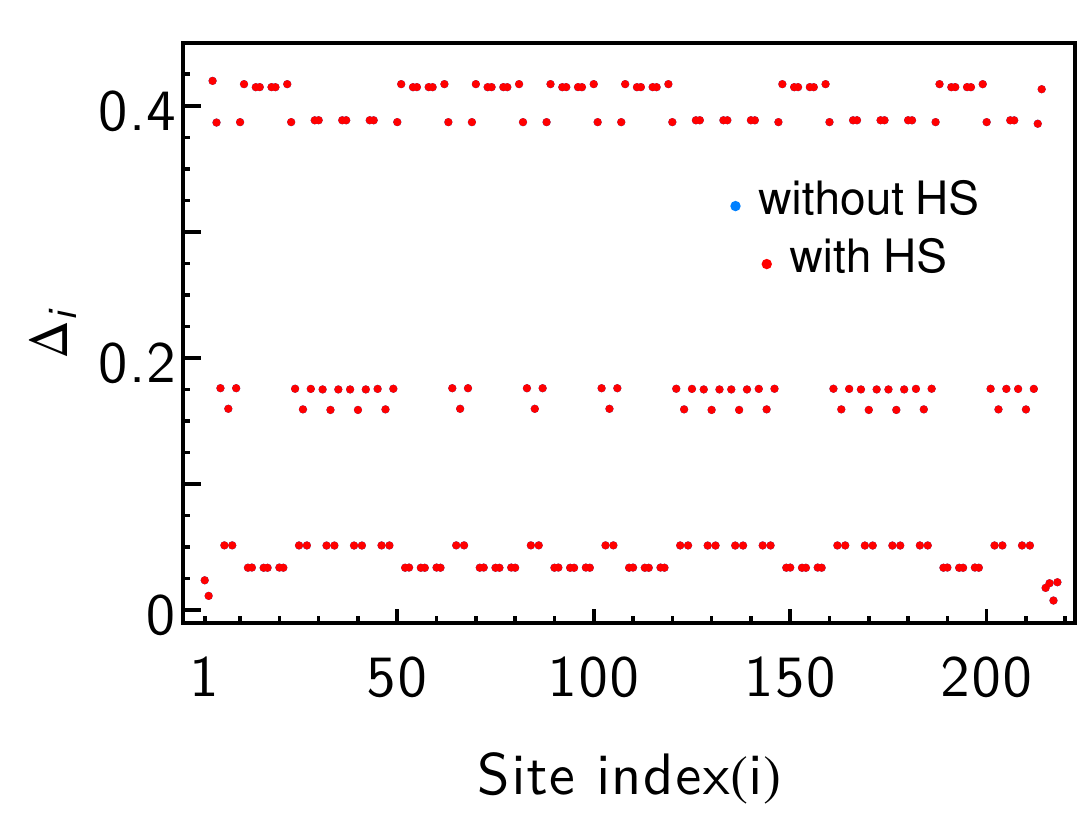}
\caption{OP profile of $1$D Nickel-mean ABS with and without Hartree shift term with $t_{A} = 1$, $t_{B} = 0.25$. The rest of the parameter values are the same as in Fig.~\ref{OI1}.}
\label{NMQCHF}
\end{figure}

Throughout this study, all results are presented without considering the Hartree shift term. In this subsection, we show the effect of the Hartree shift term with which the effective onsite potential part of the Hamiltonian becomes, 
$\epsilon_{eff} = \sum_{i,\sigma} (\epsilon+\epsilon_{i}^{HF}-\mu)c_{i,\sigma}^{\dagger}c_{i,\sigma}$. The chemical potential $\mu$ can be determined by the average electron filling~\cite{Liyu2023}:
\begin{equation}
\langle n_{e}\rangle= \frac{1}{N}\sum_{i}n_{e}(i)
\end{equation}
where,
\begin{equation}
    n_{e}(i) = 2 \sum_{n} [|u_{i}^n|^2f(E_{n})+|v_{i}^n|^2(1-f(E_{n})].
\end{equation}
At the half filling, $\langle n_{e} \rangle$ = 1. The self-consistency equation for the Hartree shift becomes: $\epsilon_{i}^{HF} = - \frac{U}{2} n_{e}(i)$~\cite{Tanaka2000}. Now, to calculate the OP self-consistently, we begin with an initial guess value of $\mu$, $\Delta_{i}$, and $\epsilon_{i}^{HF}$, and then calculate $n_{e}(i)$ with the new $\Delta_{i}$ and $\epsilon_{i}^{HF}$ using the quasiparticle energies and wavefunctions. This step is repeated until the convergence is achieved for both $\epsilon_i^{HF}$ and $\Delta_i$ with a tolerance of $10^{-5}$\cite{Chen2024a}. The effect of Hartree-shift can be minimized by a proper shifting of the chemical potential~\cite{Zhu2025, Bai2023a}. We find that after considering the Hartree shift term, there is a nominal change in the behavior of the OP as long as the Hubbard interaction is not strong and the condition on the chemical potential is maintained. In order to show the effect of the Hartree term, we show the behavior of the OP only for the Nickel-mean ABS as an example in Fig.~\ref{NMQCHF}. We observe that the OP profiles with and without the Hartree term almost overlap with each other. We also confirm the same for other other ABSs. Thus, the non-consideration of the Hartree-shift term is completely justified as long as the interaction is not strong and the chemical potential is tuned appropriately since the qualitative behavior remains unaffected, making nominal changes in the amplitudes.

\section{Effect of filling}
\label{Appendix_Filling}
\begin{figure}
\centering
\includegraphics[width=\linewidth]{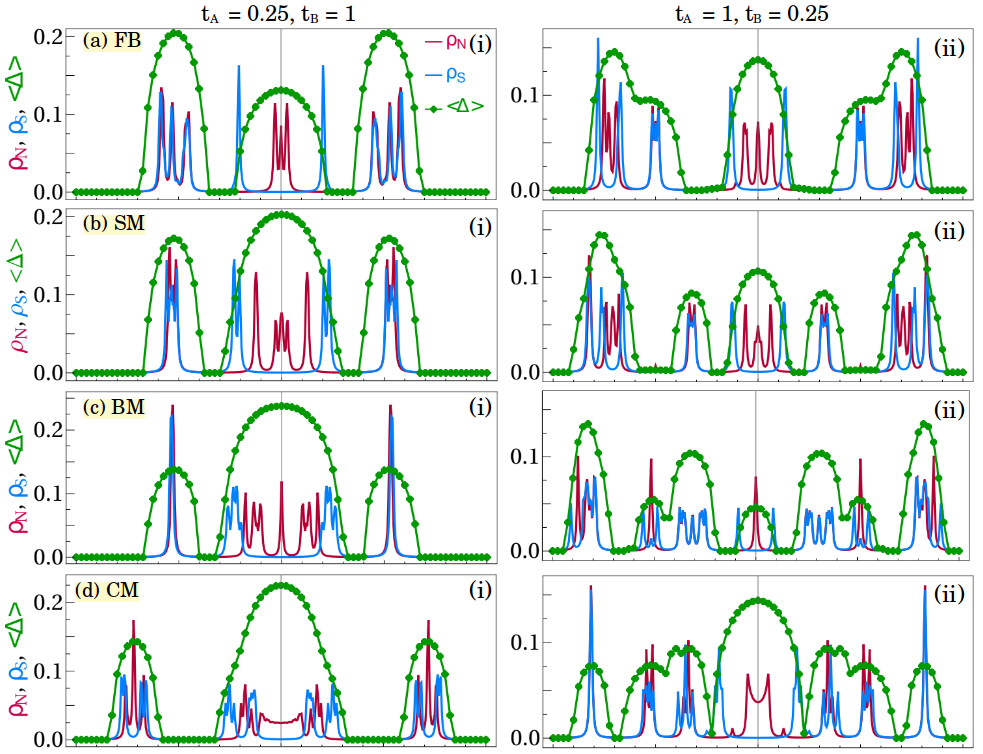}
\includegraphics[width=\linewidth]{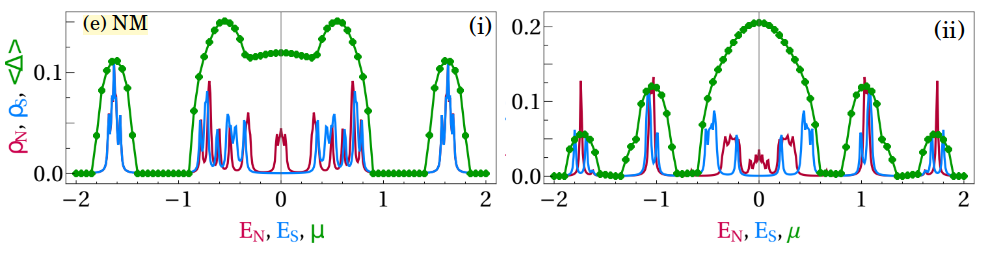}
\caption{DoS for both normal $\rho_N$ ($U = 0$) and superconducting $\rho_S$ ($U = 1$) state, and the average OP as a function of energy $E_N, E_S$ and chemical potential $\mu$, respectively, for (a-c) ABS-I and (d-e) ABS-II. The parameters are chosen as (a-e)(i) $t_A = 0.25$ and $t_B = 1$, (a-e)(ii) $t_A = 1$ and $t_B = 0.25$. The rest of the parameter values and the abbreviations are the same as in Figs.\ref{OI1} and \ref{OI2}.}
\label{mu}
\end{figure}

In the main text, we have shown all results only for the half-filling, considering $\mu = 0$. For the sake of completeness, we now present DoS profiles both for the normal state ($U = 0$), $\rho_N$, and the superconducting phase ($U=1$), $\rho_S$, and the behavior of superconducting OP with the filling in Fig.~\ref{mu}. The DoS for non-interacting ($U=0$) ABSs are fragmented as discussed in the literature~\cite{Capaz1990, Jagannathan2021, Ryu1993, Mantela2019}. As soon as $U$ is turned on, the DoS profiles show the appearance of a superconducting gap at the Fermi level, surrounded by coherence peaks on both sides. However, the widths of the superconducting gaps, even for one particular parameter regime, are not the same for all ABSs. For a given parameter regime, the superconducting gap is comparable for all members of ABS-I with modifications in the coherence peak heights. The gap structure changes for ABS-II as seen in Fig.~\ref{mu}. 

The amplitude of OP is always finite for the filling within the superconducting gap across all ABSs. If $\mu$ falls within `gapless' regions of the spectrum, the ABSs can host non-zero OP since those regions are not truly gapless; rather fragmented. This feature differs from the typical behavior of ordinary superconductors and also of unconventional nodal superconductors. When $\mu$ is tuned to well-defined gapped regions, it may or may not show non-zero OP. It will be governed by the type of gap, whether it is the spectral or the superconducting gap. Note that, the OP domes will be sharper if we scan the filling at further smaller interval. Due to the computational resource limitations, we consider the present size with an optimal scanning where the qualitative effect of the filling on the OP is clear. Comparing the OP vs filling plots for the Fibonacci with other metallic means, we see that the overall dome-like pattern is qualitatively similar except, giving rise to similar OP for other members of ABS family up to a shift in the filling for a given parameter regime. 

As discussed in the main text, all systems show a finite OP at the half-filling, i.e. $\mu = 0$, taking contributions by $\alpha$ (atomic) sites together. Among all ABSs considered here, the Bronze-mean shows the highest average OP amplitude at half filling since it contains the maximum number of $\alpha$ sites. However, beyond half-filling, the OP is higher mostly at the $\gamma$ (molecular) sites. Given that the number of this type of site is higher for the Fibonacci sequence, this ABS supports higher average OP amplitudes beyond half-filling. Similarly, the behavior of the average OP with $\mu$ for ABS-II can be explained. The DoS as well as the OP values strongly depend on the parameter regime. When we tune the parameter values, the normal state gaps as well as the superconducting gap widths change, resulting in a change in the OP profile. 
\begin{figure}
\centering
\includegraphics[width=\linewidth]{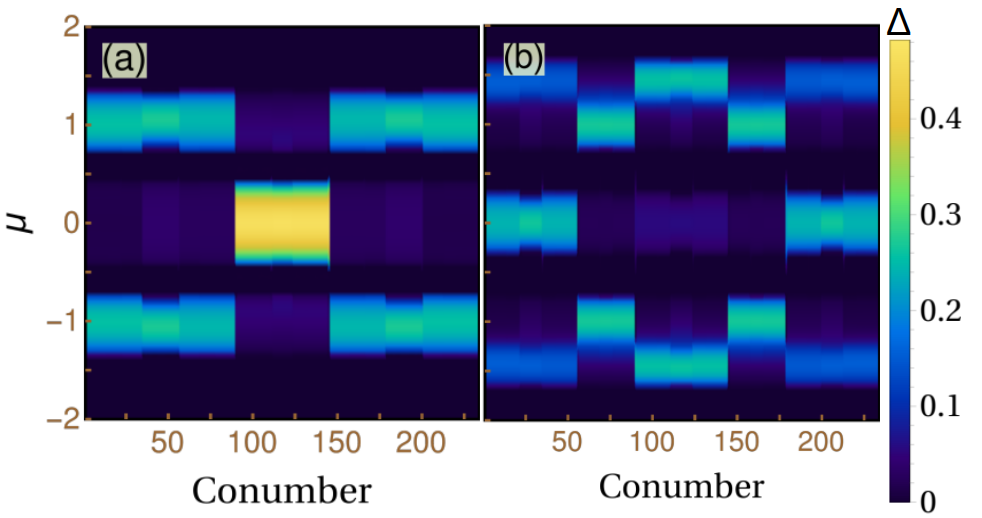}
\caption{Distribution of OP with chemical potential $\mu$ over conumber space for Fibonacci ABS. The parameters are chosen as (a) $t_A = 0.25$ and $t_B = 1$, (b) $t_A = 1$ and $t_B = 0.25$. The rest of the parameters are fixed as in Fig.~\ref{OI1}.}
\label{conumber_mu}
\end{figure}

The effect of filling can also be understood by using the conumber space analysis. Figure~\ref{conumber_mu} demonstrates the behavior of the average OP with $\mu$ over the conumber space. At half filling, for $t_A<t_B$, the finite OP appears mainly at $\alpha$ sites and beyond half-filling it is mainly hosted by $\gamma$ sites as shown in Fig.~\ref{conumber_mu}(a). It exactly matches the previously discussed spectral distribution (see Fig.~\ref{Conumber}(c)). In the reverse parameter regime, $\alpha$ sites no longer support the higher OP at half filling due to their low spectral weight. Instead, higher OP is observed at $\gamma$ sites, which have higher spectral weight as confirmed from Fig.~\ref{conumber_mu}(b) and Fig.~\ref{Conumber}(d). The variations in the dome heights in Fig.~\ref{mu} can be understood from the density plot of Fig.~\ref{conumber_mu} too. Thus, the spectral distribution and the variation of the average OP with $\mu$ over the conumber space relate to each other and give confirmation of the OP behavior for all fillings.
\section{Convergence to periodic and non-interacting limit}
\label{Appendix_periodic_limit}

\begin{figure}
\centering
\includegraphics[width=\linewidth]{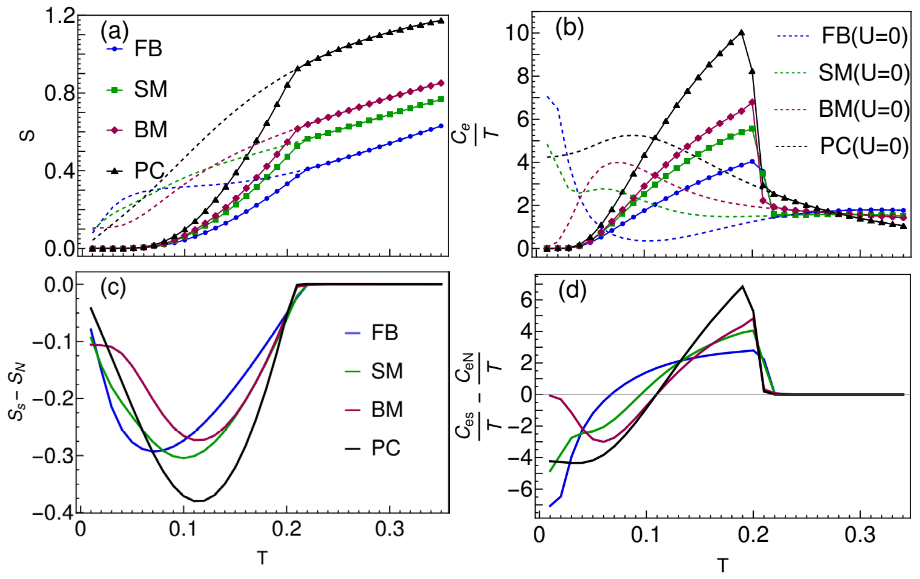}
\caption{(a) Entropy and (b) electronic specific heat as a function of temperature for ABS-I with $t_{A} = 0.25$, $t_{B} = 1$, and a periodic chain with $t_A = t_B = 0.25$ in the absence and presence of Hubbard term $U$. The difference in (c) entropy and (d) electronic specific heat between superconducting state $(U =1)$ and normal state $(U = 0)$ are also displayed. The rest of the parameter values are the same as in Fig.~\ref{OI1} and the abbreviations are the same as in Fig.~\ref{tatb} of the main text.}
\label{ENCEPC}
\end{figure}

To understand how the thermodynamic quantities evolve when moving from the Fibonacci to the higher metallic mean sequence, we plot their temperature dependence in the presence of the Hubbard term $U$ and compare with the results of the periodic chain. In Fig.~\ref{ENCEPC}, we show results only for $t_A < t_B$ to avoid repetition of the similar analysis in the other parameter regime. To understand the jump in the electronic specific heat, we also calculate it for the normal-state without any attractive Hubbard interaction, i.e., at $U = 0$ for the ABS-I family, along with that of the periodic chain, for the sake of comparison and present them in Fig.~\ref{ENCEPC}(b). It is clearly visible that the magnitudes of the both the quantities are suppressed in ABSs compared to the periodic system which can be explained from the comparative analysis of the OP below the transition temperature. They show typical BCS behaviors but the height of the jump of the specific heat is suppressed for the Fibonacci chain but increases as we move to Silver-mean and Bronze-mean ABS. It happens since as we move towards the higher metallic mean sequences, it should approach the periodic limit. Above the transition point, the specific heat curves finite $U$ converge to the metallic limit, as shown in Fig.~\ref{ENCEPC}(b). 
For $U = 0$, Cooper pairs cannot form, leading to a vanishing OP irrespective of the temperature. Consequently, no sudden drop in the electronic specific heat is observed throughout the whole temperature regime because of the absence of the phase transition. In contrast, when $U$ is finite, a nonzero OP emerges at low temperatures, followed by a sharp drop in specific heat at the transition temperature, beyond which the OP vanishes. 

For a clear comparison, we compute the difference between the superconducting and normal state entropy, and electronic specific heat as a function of temperature, as depicted in Fig.~\ref{ENCEPC}(c-d). Note that the difference between the entropies in the superconducting and normal phases is negative below the transition temperature since the superconducting states containing the Cooper pairs are ordered phases exhibiting lower entropies for all ABSs. However, above the transition point, there are no Cooper pairs; systems move to the normal phase, showing no difference in entropy. It clearly shows that differences in the entropy ($S_S-S_N$) for various ABSs are comparable to each other and not much smaller than the periodic counterpart. Moreover, the electronic specific heat difference helps quantify the transition point and the jumps. There are some anomalies which corroborate the discussions of the main text.

\section{Some additional results for ABS-III}
\label{Appendix_ABSIII}

\begin{figure}
\centering
\includegraphics[width=\linewidth]{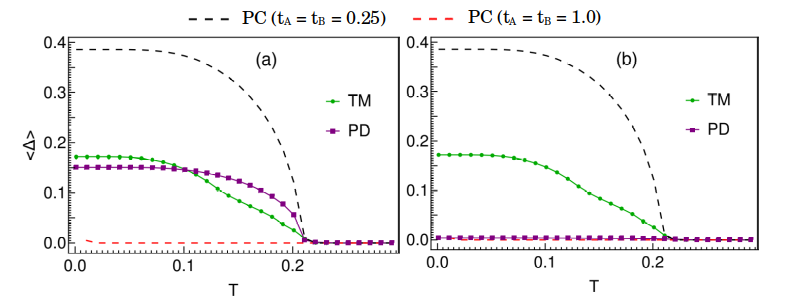}
\caption{Behavior of the average OP with temperature for Thue-Morse (TM) and Period-doubling (PD) of ABS-III with (a) $t_{A} = 0.25$, $t_{B} = 1$ and (b) $t_{A} = 1$, $t_{B} = 0.25$. The rest of the parameter values and abbreviations are maintained as in Fig.~\ref{OI1}.}
\label{OT_TM_PD}
\end{figure}

In this section, we present some additional results for ABS-III. The temperature dependence of the average OP for Thue-Morse and period-doubling ABSs are displayed in Fig.~\ref{OT_TM_PD} for both parameter regimes. For $t_{A} < t_{B}$ (see Fig.~\ref{OT_TM_PD}(a)), the behavior of the OP is very similar to that for the QPS since both these two ABSs contain significant number of $\alpha$ sites that cause a higher values of OP. It deviates from the typical BCS-like profile for the Thue-Morse lattice. In the period doubling, the number of $t_A$ bonds are much higher than that in the Thue-Morse. This results in higher number of $\alpha$ sites which effectively leads towards the behaviors of the higher metallic mean.  

We noticed that the temperature dependence of the average OP is not identical for the Thue–Morse and Period-doubling ABS. The Period-doubling ABS exhibits the expected BCS-like behavior: the OP landscape consists of only a single non-zero branch, contributed solely by the $\alpha$ sites. As the temperature increases, electrons are thermally excited, making Cooper pair formation progressively less probable. Consequently, the OP magnitude on all $\alpha$ sites decreases, and at the transition temperature, the entire OP branch vanishes. In contrast, the Thue–Morse ABS shows a distinctly different behavior due to the presence of two non-zero OP branches. With increasing temperature, the lower non-zero branch, originating from the $\gamma$ sites of the $\gamma$–$\beta$–$\gamma$ cluster, disappears first. This gives rise to a slight curvature in the temperature dependence of the average OP. The higher non-zero branch, formed by the $\alpha$ sites, continues to survive beyond this point but gradually decreases in magnitude with further temperature increase, eventually vanishing at the transition temperature. Note that, the local suppression around $T=0.15$ will be decreasing with the increase in system size. We kept this system size to maintain consistency with other systems.

Conversely, when we choose the opposite parameter regime as $t_{A} > t_{B}$, the average OP versus temperature curve for the Thue-Morse ABS  maintains a similar profile to the previous parameter regime. It shows a small curvature due to the presence of two non-zero OP branches (see main text). The explanation follows the similar reasoning as discussed earlier for the previous parameter regime. The difference here is that the lower non-zero branch originates from the $\gamma$ sites of the $\gamma$–$\alpha$–$\gamma$ cluster, while the higher non-zero branch is formed by all the $\beta$ sites. As the temperature increases, the lower non-zero branch disappears first, and finally, at the transition point, the higher non-zero branch also vanishes. On the other hand, the variation of the average OP is very small for the Period-doubling case, where the amplitude of OP goes to zero due to the absence of any $\beta$ site or $\gamma$-$\alpha$-$\gamma$ cluster at any finite temperature. Now, the $\beta$ and $\gamma$ sites (in $\gamma$-$\alpha$-$\gamma$ cluster) play the main role in containing the higher OP values. As both structural units are present in the Thue-Morse ABS, it has a high average OP at low temperature regime as shown in Fig.~\ref{OT_TM_PD}(b).
\begin{figure}
\centering
\includegraphics[width=\linewidth]{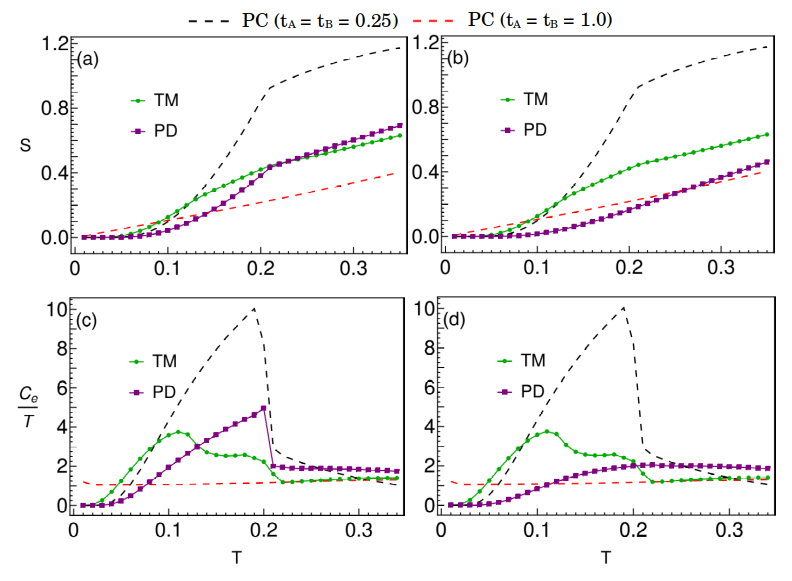}
\caption{(a,b) Entropy and (c,d) electronic specific heat as a function of temperature for the Thu-Morse (TM) and period-doubling (PD) ABS-III with (a,c) $t_{A} = 0.25$, $t_{B} = 1$ and (b,d) $t_{A} = 1$, $t_{B} = 0.25$. The rest of the parameter values and abbreviations are maintained as in Fig.~\ref{OI1}.}
\label{EN-Ce-TM_PD}
\end{figure}

The variation of the entropy and electronic specific heat with temperature for the two members of ABS-III is shown in Fig.~\ref{EN-Ce-TM_PD}. Interestingly, the variation of the entropy and the specific heat with the temperature is different for the two systems depending on the parameter regime. We avoid repetition of the explanation when the behaviors of these two systems follow our analysis in the main text. We summarize and analyze only the noticeable changes in the behaviors of these two ABS-III here. For $t_A < t_B$, the entropy behaviors are quite similar to what we found for ABS-I and ABS-II. For the same parameter regime, the Period-doubling chain shows a sharp fall in the $C_e$ profile at the transition point, while the $C_e$ profile for the Thue-Morse chain takes a curvature. For the reverse parameter regime, the entropy for the Period-doubling chain shows slowly increasing behavior and also the $c_e$ profile does not show any sharp fall, which happens due to the suppression in OP amplitudes. Interestingly, the Thue-Morse ABS shows the unusual behavior for both parameter regimes, as expected due to the branching of the OP following the discussions in the main text. The electronic specific heat $C_e$ exhibits two successive drops at different temperatures. The first drop is linked to the disappearance of the lower non-zero OP branch, while the second, more pronounced drop corresponds to the eventual vanishing of the higher non-zero OP branch as temperature increases. This two-step feature is a clear manifestation of the multi-branch OP landscape characteristic of the Thue–Morse ABS.
For better understanding, we also compare the temperature dependence OP and thermodynamics quantities with the periodic limits. Although the amplitude of OP differ, the transition temperature appears nearly same point for the Thue-Morse ABS in both parameter regimes. This is also valid for Periodic-doubling ABS for $t_A<t_B$ regime.
\begin{figure}
\centering
\includegraphics[width=0.6\linewidth]{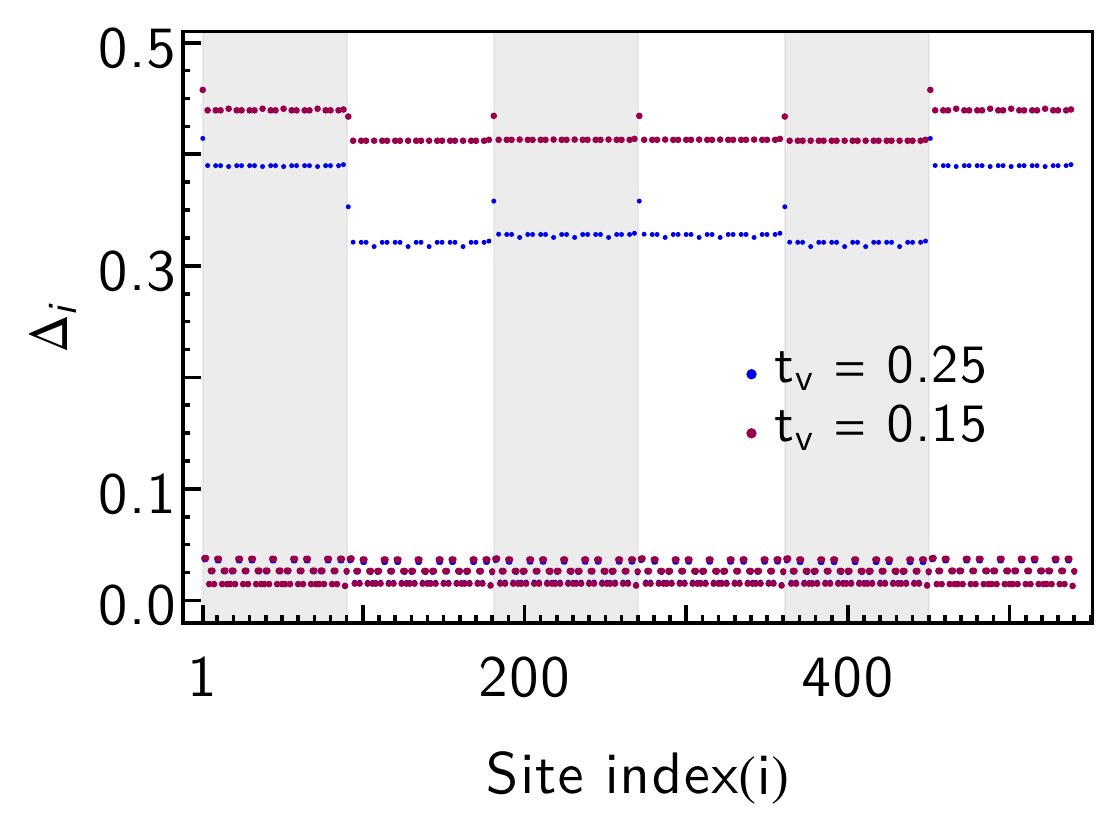}
\caption{Distribution of the OP in a $2$D Fibonacci QPS with $M\times N=6\times 90$ sites, $t_{A} = 0.25$, $t_{B} = 1$ for different $t_v$ values. We mark odd M ($M =1, 3, 5$) by gray shades. The rest of the parameter values are the same as in Fig.~\ref{OI1}.}
\label{OItv}
\end{figure}

\section{Some additional results for $2$D ABSs }
\label{Appendix_2D}

In this section, we present a few more results on $2$D ABS to support our conclusion of the present work.

In order to show the effect of the vertical hopping, we show the results of Fibonacci QPS for different vertical hoppings: (i) $t_v = 0.25$ (blue points, exactly identical to Fig.~\ref{OI_2D_FIBONACCI1}(a)) and (ii) $t_v = 0.15$ (pink points) in Fig.~\ref{OItv}. The formation of $2$D ABS allows three different hopping amplitudes, i.e., $t_A$, $t_B$, and $t_v$. It is evident that for the smaller vertical hopping strength, the system consistently exhibits higher OP magnitudes. This finding is equally applicable to all other $2$D ABSs including QPSs for both parameter regimes. Thus, the distribution of OP across the lattice i.e., the branching phenomena determined by the sequence remains invariant qualitatively. It happens since we have the quasiperiodicity only along the $x$-axis, the vertical hoppings remain the same. 
\begin{figure}[ht!]
\centering
\includegraphics[width=\linewidth]{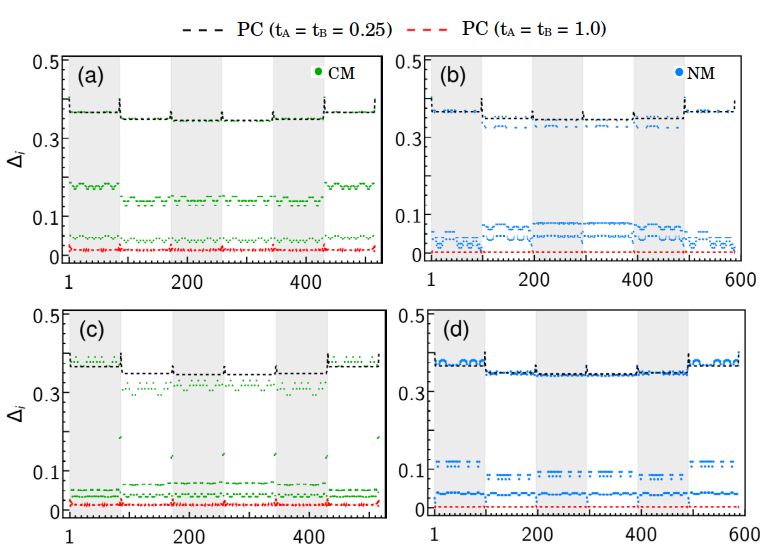}
\includegraphics[width=\linewidth]{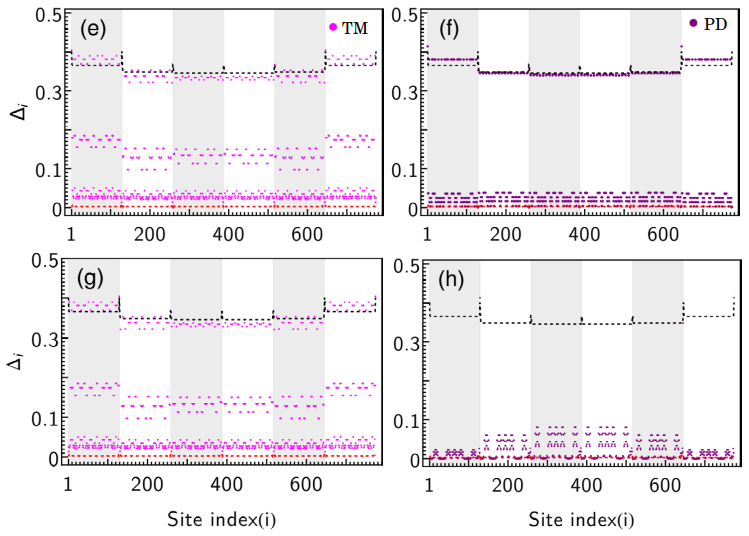}
\caption{Distribution of OP over sites for $2$D with $M\times N$ sites (a,c) Copper-mean ($M=6,N=86$), (b,d) Nickel-mean ($M=6,N=98$), (e,g) Thue-Morse ($M=6,N=129$) and (f,h) Period-doubling ($M=6,N=129$) ABSs. The hopping amplitudes are taken as (a,b,e,f) $t_{A} = 0.25$, $t_{B} = 1$ and (c,d,g,h) $t_{A} = 1$, $t_{B} = 0.25$. The vertical hopping strength is set at $t_{v} = 0.25$. We mark odd M ($M =1, 3, 5$) by gray shades. The rest of the parameter values are set as the same in Fig.~\ref{OI1}. The abbreviations are the same as in Fig.~\ref{OI2} and \ref{OI-TM-PD}.}
\label{OI_2D_FIBONACCI2}
\end{figure}

The behavior of OP profiles for $2$D ABS-II is demonstrated in Fig.~\ref{OI_2D_FIBONACCI2}(a-d). We consider $G_7$-th Copper-mean ABS containing $516$ sites and $G_6$-th Nickel-mean  ABS containing $588$ sites. Within the parameter regime $t_A<t_B$, the overall OP distribution breaks into three and two branches for $2$D Copper-mean and Nickel-mean ABS, respectively (see Fig.~\ref{OI_2D_FIBONACCI2} (a-b)). For the Copper-mean ABS, in this parameter regime, the non-zero OP values are distributed across two distinct branches. The $\alpha$ sites and $\gamma$ sites of the $\gamma$–$\beta$–$\gamma$ cluster are responsible for forming the higher and lower non-zero OP branches, respectively. In contrast, Cooper pair formation at the $\beta$ sites is much less probable. Consequently, the OP magnitude at these sites tends to vanish, giving rise to the zero branch in the OP profile (see Fig.~\ref{OI_2D_FIBONACCI2}(a)). For the Nickel-mean ABS, the OP profile also exhibits two distinct branches. Here, the higher OP branch originates from the $\alpha$ sites, while the $\beta$ and $\gamma$ sites carry lower OP magnitudes, collectively forming the lower branch (see Fig.~\ref{OI_2D_FIBONACCI2}(b)). Similarly, the OP profiles at reverse regime are depicted in 
Fig.~\ref{OI_2D_FIBONACCI2}(c-d). In the parameter regime $t_A > t_B$, the higher OP values are located at the $\beta$ sites. For the Copper-mean ABS, the OP profile is organized into two main branches: the upper branch, formed entirely by the $\beta$ sites, and the lower branch, contributed by the $\alpha$ and $\gamma$ sites (see Fig.~\ref{OI_2D_FIBONACCI2}(c)). In the case of the Nickel-mean ABS, the situation is similar, with the higher non-zero branch again originating from the $\beta$ sites. In addition, the $\gamma$ sites and the central $\alpha$ sites within the $\gamma$–$\alpha$–$\alpha$–$\alpha$–$\gamma$ cluster also support finite OP values (though smaller compared to the $\beta$ sites), giving rise to a secondary lower non-zero branch. Finally, the lowest branch in the OP profile emerges due to the very small OP values on the remaining $\alpha$ and $\gamma$ sites (see Fig.~\ref{OI_2D_FIBONACCI2}(d)). So, the overall behavior of the OP landscape is preserved, not affected by the extension in the dimension
\begin{figure}
\centering
\includegraphics[width=0.7\linewidth]{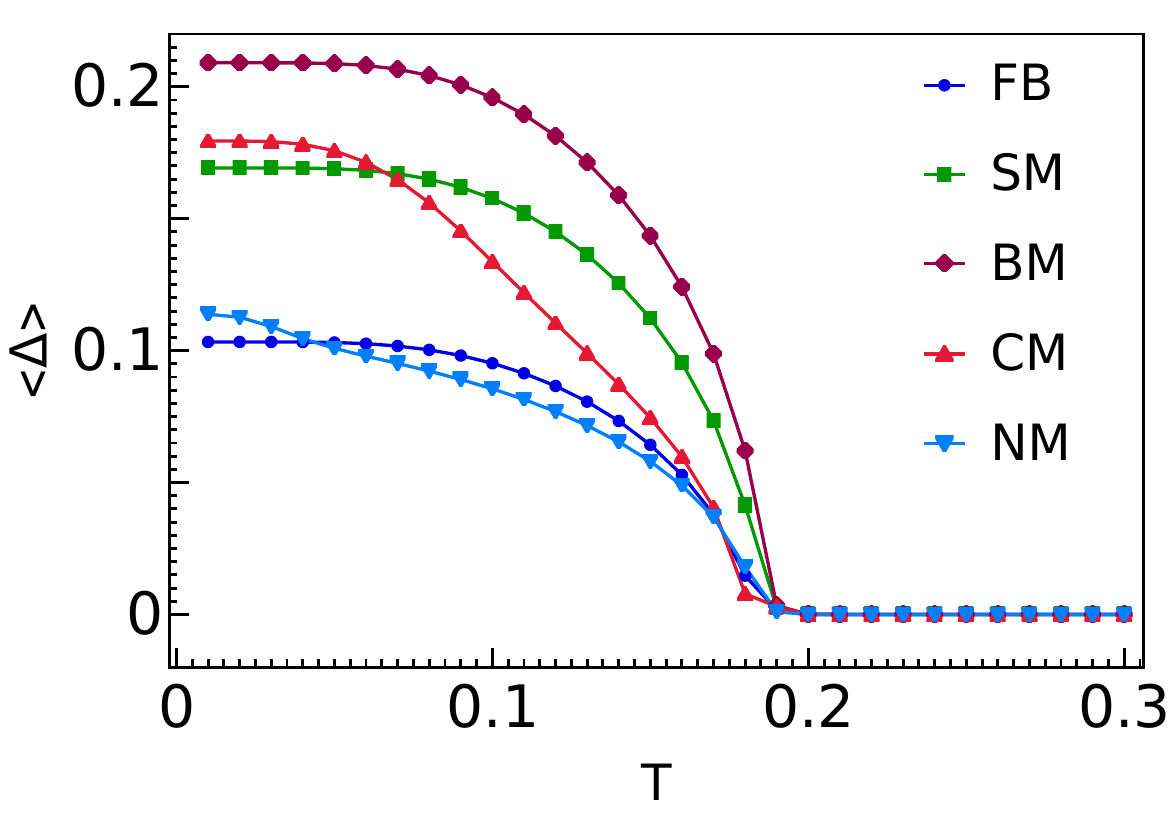}
\caption{Variation of average OP with temperature for $2$D ABS-I and ABS-II. The parameters are chosen as $t_{A} = 0.25$, $t_{B} = 1$, $t_{v} = 0.25$. The rest of the parameter values are maintained as in Fig.~\ref{OI1}, and the abbreviations are the same as in Fig.~\ref{tatb}. The dimensions of the systems are the same as in Fig.~\ref{OI_2D_FIBONACCI1} and Fig.~\ref{OI_2D_FIBONACCI2}.}
\label{OT_2D}
\end{figure}

We extend our study to $2$D ABS-III too for the completeness of the present work. Here also, the $2$D OP profiles show similar behavior compared to $1$D cases, indicating that the underlying mechanism persists across dimensions. We consider $2$D the Thue-Morse and period-doubling ABSs in their $G_8$-th generation, containing $774$ sites. In $2$D Thue-Morse ABS, the OP profiles are distributed over three branches for both parameter regimes as shown in Fig.~\ref{OI_2D_FIBONACCI2}(e-g). In the parameter regime $t_A < t_B$, the higher OP branch originates from the $\alpha$ sites. In addition, the $\gamma$ sites belonging to the $\gamma$–$\beta$–$\gamma$ cluster possess finite OP values and are responsible for forming a middle branch. The lower branch, on the other hand, is formed by the $\beta$ sites together with the remaining $\gamma$ sites that are not part of the $\gamma$–$\beta$–$\gamma$ cluster (see Fig.~\ref{OI_2D_FIBONACCI2}(e)). 

In the opposite regime, $t_A > t_B$, the situation is completely reversed. Here, the $\beta$ sites host the largest OP values, giving rise to the upper branch. The middle branch now arises from the $\gamma$ sites of the $\gamma$–$\alpha$–$\gamma$ cluster, while the lowest branch is formed by the $\alpha$ sites along with the other $\gamma$ sites (see Fig.~\ref{OI_2D_FIBONACCI2}(g)). 
\begin{figure}
\centering
\includegraphics[width=0.498\linewidth]{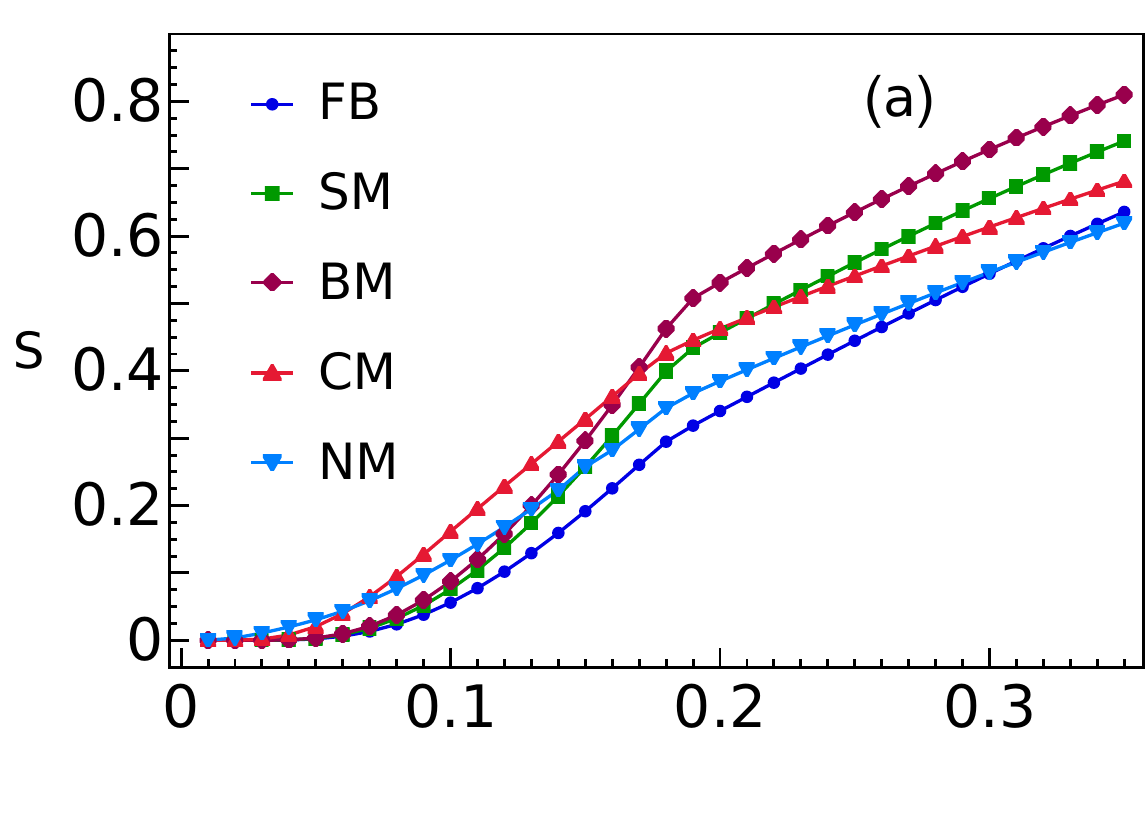}
\includegraphics[width=0.48\linewidth]{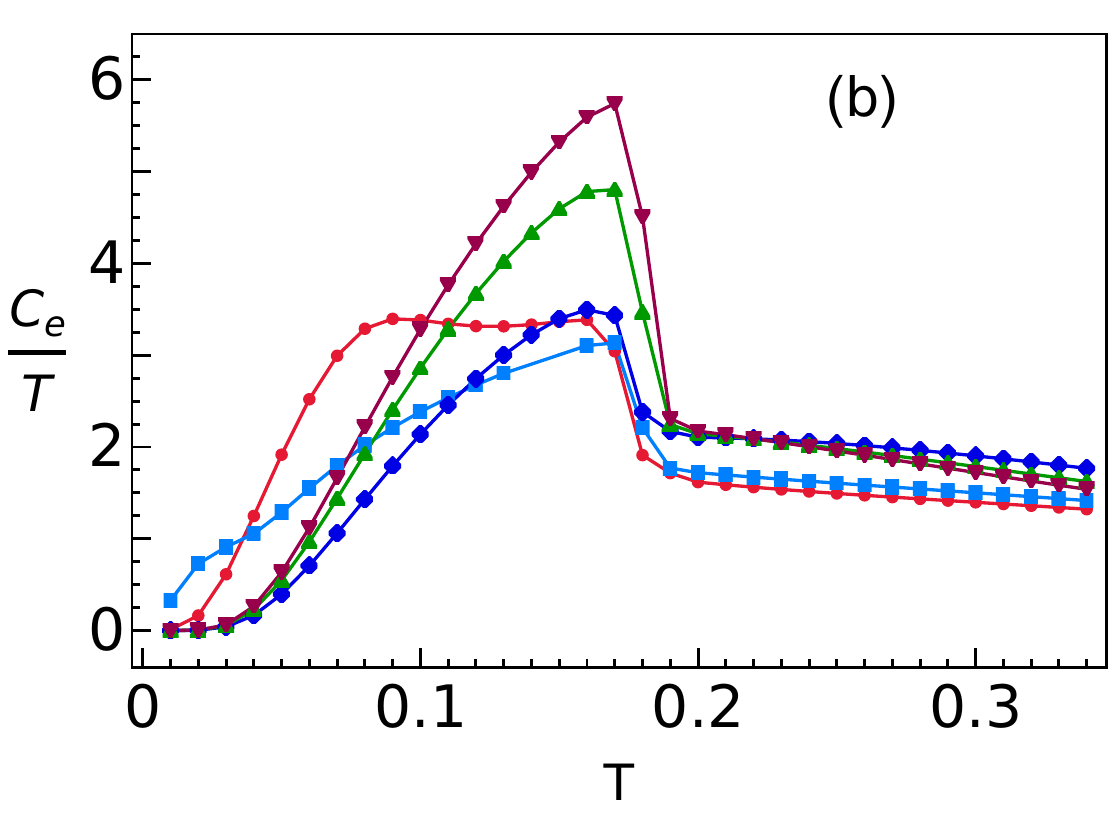}
\caption{Variation of (a) entropy and (b) electronic specific heat with temperature for $2$D ABS-I and ABS-II. The parameters are chosen as $t_{A} = 0.25$, $t_{B} = 1$, $t_{v} = 0.25$. The rest of the parameter values are maintained as in Fig.~\ref{OI1} and the abbreviations are the same as in Fig.~\ref{tatb}. The dimensions of the systems are the same as in Fig.~\ref{OI_2D_FIBONACCI1} and Fig.~\ref{OI_2D_FIBONACCI2}.}
\label{ENCV_2D}
\end{figure}
Moreover, in $2$D period-doubling ABS, the OP mainly shows two branches at $t_A<t_{B}$ (see Fig.~\ref{OI_2D_FIBONACCI2} (f)). The higher branch is made by all $\alpha$ sites, and the lower branch contains the OP values of all $\gamma$ sites. Conversely, in a reverse regime, the OP distribution of $2$D period-doubling ABS becomes suppressed as depicted in Fig.~\ref{OI_2D_FIBONACCI2}(h). This behavior is expected, since in this parameter regime the larger OP magnitudes are associated with the $\beta$ sites or with the $\gamma$ sites belonging to the $\gamma$–$\alpha$–$\gamma$ cluster. However, such site configurations do not occur in the Period-doubling ABS because of the intrinsic sequential frustration built into its structure. Consequently, the system lacks the favorable local environments that support higher OP values, and the overall OP landscape is significantly suppressed in this regime. Apart from the distribution of the OP over $2$D ABS, the temperature dependence of the average OP also shows behavior similar to $1$D counterparts. The average OP variation with temperature for $2$D ABS-I (QPS) and ABS-II family is shown in Fig.~\ref{OT_2D} within the parameter regime $t_A<t_B$. Here also at low temperature regime, the Bronze-mean QPS exhibits the highest average OP magnitude as seen in the  $1$D Bronze-mean QPS, similar to the behavior seen in $1$D cases (see Fig~\ref{OT}(a)). For the contrast parameter regime, the behaviors of the average OP in these $2$D ABSs are also very similar to that in $1$D counterparts and can be analyzed in exactly the same way.

For completeness, we also compute the thermodynamic quantities for the $2$D ABS-I and ABS-II and present them in Fig.~\ref{ENCV_2D}. We only show the results for the parameter regime $t_A < t_B$ to avoid repetition. It will also not affect the main message of the present work. The temperature dependence of the entropy, shown in Fig.~\ref{ENCV_2D}(a), exhibits a similar trend to that observed in the $1$D case (see Fig.~\ref{En_Ce}(a)). Furthermore, Fig.~\ref{ENCV_2D}(b) illustrates the temperature variation of the electronic specific heat, which clearly indicates a superconducting-to-normal phase transition marked by a sharp drop at the critical temperature. This behavior closely resembles the results obtained for the one-dimensional systems, as displayed in Fig.~\ref{En_Ce}(c). The origin of such features, including the unusual behavior of the Copper-mean ABS, has already been discussed in the context of the $1$D ABSs, and the same explanation holds for the $2$D systems. It is worth noting that small discrepancies and fluctuations appear in the Copper-mean and Nickel-mean ABSs, which can be attributed to finite-size effects. These deviations can be minimized by increasing the system size with higher generation; however, this is computationally challenging due to limited numerical resources. 

\bibliography{ref}{}
\end{document}